\documentclass[
 reprint,
superscriptaddress,
 amsmath,amssymb,
 aps,
]{revtex4-2}
\usepackage{CJK}
\usepackage{graphicx,color}
\usepackage{dcolumn}
\usepackage{bm}
\usepackage[mathscr]{eucal}
\usepackage{pxfonts}
\usepackage{graphicx}
\usepackage{ulem}
\usepackage[urlcolor=blue,colorlinks=true,citecolor=blue,breaklinks]{hyperref}
\usepackage{braket}

\usepackage{subfigure}

\usepackage{epsfig}
\usepackage[]{natbib}
\usepackage{xspace}
\usepackage{color}
\usepackage{bbold}
\usepackage{tabu}
\usepackage{mathtools}

\begin{document}

\title{First Hitting Times on a Quantum Computer: \\Tracking vs. Local Monitoring, Topological Effects, and Dark States}
\author{Qingyuan Wang}
     \email{qingwqy@gmail.com}
\affiliation{Department of Physics, Institute of Nanotechnology and Advanced Materials, Bar Ilan University, Ramat-Gan
52900, Israel}
\author{Silin Ren}
\affiliation{Department of Physics, Institute of Nanotechnology and Advanced Materials, Bar Ilan University, Ramat-Gan
52900, Israel}
\author{Ruoyu Yin}
 \email{yinruoy@biu.ac.il}
\affiliation{Department of Physics, Institute of Nanotechnology and Advanced Materials, Bar Ilan University, Ramat-Gan
52900, Israel}
 \author{Klaus Ziegler}
 \affiliation{
 Institut f\"ur Physik, Universit\"at Augsburg, D-86135 Augsburg, Germany%
}
\author{Eli Barkai }
\affiliation{Department of Physics, Institute of Nanotechnology and Advanced Materials, Bar Ilan University, Ramat-Gan
52900, Israel}
\author{Sabine Tornow}

  \email{Sabine.Tornow@unibw.de}
 \affiliation{Research Institute CODE, University of the Bundeswehr Munich, D-81739 Munich, Germany} 
 
\begin{abstract}
We investigate a quantum walk on a ring represented by a directed triangle graph with complex edge
weights and monitored at a constant rate until the quantum walker is detected.
To this end, the first hitting time statistics is
recorded using unitary dynamics interspersed stroboscopically
by measurements, which is implemented on IBM quantum computers with a midcircuit readout option.
Unlike classical hitting times, the statistical aspect of the problem
depends on the way we construct the measured path, an effect that we quantify experimentally. 
First, we experimentally verify the theoretical prediction that the mean return time to a target state is quantized, with abrupt discontinuities found for specific
sampling times and other control parameters, which has a well-known topological interpretation.
Second, depending on the initial state, system parameters, and measurement protocol, the detection probability can be less than one or even zero, which is related to dark-state physics.
Both, return-time quantization and the appearance of the dark states are related to degeneracies in the eigenvalues of the unitary time evolution operator.
 We conclude that, for the IBM quantum computer under study,
 the first hitting times of monitored quantum walks are resilient to noise. Yet, a finite number of measurements leads to broadening effects, which modify the topological quantization and chiral effects of the asymptotic theory with an infinite number of measurements.
 Our results point the way for the development of novel quantum walk algorithms that exploit measurement-induced effects on quantum computers.
\end{abstract}
\maketitle

\section{Introduction}

The option of midcircuit readout of qubits on state-of-the-art quantum computers \cite{Motta} opens opportunities to test the dynamics and statistics of monitored quantum dynamics.
The repeated monitoring at predetermined times yields
a string of measurement outputs.
These can be viewed as a stochastic trajectory varying in time which
depends on many factors, including the initial state,
the dynamics of the measurement-free process, or
the chosen time intervals between measurements.
Given a stochastic trajectory, the first quantum
 hitting time, 
defined
as the time taken for a monitored process or a signal to reach a specific level or
target for the first time, has attracted considerable attention 
\cite{krovi06a,Todd2006,Todd2008,Gruenbaum2013,Grunbaum2014,Dhar2015,Dhar_2015,Friedman2017,thiel18,Yin2019,Thiel2020,Modak,kulkarni2023detection,Wang,Meng,KesslerRand,Stina,Laneve2023hittingtimesgeneral,Magniezhittingtimesrandomquantum}.
The classical counterpart, known as the first passage time
problem, was the subject
of a very large body of work; see
\cite{Redner2001,REDNER2023128545,ralf2014first} and references therein.

 Consider, as an example, a classical random walker on a lattice of dimension
$d$ \cite{Redner2001,Polya}. The classical walker
starts at the origin, and the first hitting time is the time it takes
the classical walker to reach some other vertex. The basic questions
are: Will the classical walker reach the target after in principle an infinite
number of steps? What is the distribution of the first passage
times to the target state?
The mean first hitting time is a widely used quantifier of the
process. In finite systems, excluding ergodicity breaking, simple
random walks are recurrent, and hence the classical walker is detected with probability one, i.e., $P_{{\rm det}}=1$, in any dimension. For infinite systems, the process
is non-recurrent in dimension $3$ and above and hence $P_{\rm det} < 1$. For the purpose of search, the processes are diffusive and hence non-efficient
in the sense that paths resample previously visited locations many times.

For quantum walks, which are monitored by repeated projective measurements, 
 the situation is vastly different,
and here we mention four such aspects.
\begin{itemize}
\item[1]
 {\it Constructive and destructive interference}.
Quantum walks, for example, tight-binding models (studied here), serve as a benchmark for quantum search \cite{doi:10.1080/00107151031000110776,Kempe2003}. These walks may exhibit a quantum speedup for the hitting time, which can be exponential for specific initial states
and highly symmetric graphs \cite{krovi06a}. 
However, in some other cases, the quantum search can perform poorly due to destructive interference, and the latter problem can be avoided in
principle using specially designed graphs \cite{PhysRevResearch5023141} or with a restart strategy \cite{PhysRevLett.130.050802}.
\item[2]
{\it Path definition matters}.
When we obtain quantum trajectories through repeated projective
 measurements,
the statistics of the first detection time will depend on the observation
scheme with
which the path is constructed. 
Here, we address this issue using IBM quantum computers. 
For that purpose, the first hitting time for tracking and
for localized (on-site) measurements at the target state is defined and studied.
We will also highlight when these protocols exhibit a classical, in the sense of classical random walk behaviors, and when do they exhibit quantum features.
\item[3]
{\it Topological effects}.
A case study is the return or recurrence problem in finite systems. The return problem
addresses the time it takes for a process to return to its initial state \cite{PhysRevLett100020501,Gruenbaum2013,Grunbaum2014,nitsche18,Yin2019,Wang}.
In the classical domain, the mean recurrence time is described by Kac's lemma \cite{Kac}. 
This gives the mean return time in terms of the
steady-state measure (see below). 
In a pioneering work, Gr\"unbaum {\it et al.} \cite{Gruenbaum2013,Grunbaum2014}
 extended this result to the quantum domain.
They showed that the mean number of measurements until the detection
of the quantum walker in its initial state is quantized. This is connected
to a topological effect. Mathematically, the mean return time is related
to a winding number of the generating function of the first detection
statistics.
One of the most striking differences between classical and 
quantum walks is a phase in the latter.
Although the phase is usually not directly observable, its
properties can have a significant impact on observable quantities.
Here, the mean return time is topologically protected; however, sharp transitions are found
for special choices of the sampling time. 
It turns out that the mean return time 
number is directly linked to the winding number, which 
implies a quantization of the latter \cite{Gruenbaum2013,Grunbaum2014}. Moreover, the winding 
number of the return problem is equal to the dimensionality
of the available Hilbert space.
These results are valid for local measurements at the target state, when the information about the trajectory between measurements is unknown.

The authors of Ref.~\cite{Didi} studied a different protocol, when the trajectory is
defined with a tracking protocol. 
In this case, the position of the quantum walker is recorded along its path until its first detection in the target state.
Again, the mean return time
is quantized, being larger than or equal to the one found in Ref. \cite{Gruenbaum2013,Grunbaum2014} using local monitoring. When monitoring obtains the information about the trajectory of the quantum walker, we destroy the phase, and hence for most sampling rates the behavior we find experimentally below is classical. However, for specific sampling rates, quantum feature are shown to emerge. In contrast, the local measurement protocol investigated here exhibits robust quantum characteristics that are independent of the sampling rate.

\item[4]
{\it Zeno physics}. In quantum mechanics, sampling cannot be done too fast, to avoid freezing of the wave packet in its initial state \cite{Misra1977}. 
Thus, the data acquisition rate is an important parameter controlling the statistics
of the hitting time process in both protocols.
\end{itemize}

The primary objective of this work is to enhance our understanding of the aforementioned concepts by investigating the return time problem and dark states
on a quantum processor (IBM {\it Sherbrooke}). Specifically, our objective is to explore the
differences between monitoring one local site and monitoring all sites.
 The fundamental questions are how the two monitoring methods affect return times and detection probabilities on a quantum computer, whether they are affected by noise, and how the known properties derived by the asymptotic theory are changed due to a finite number of measurements. 
 First, we discover that IBM {\it Sherbrooke} is a well suited testbed to confirm the stochastic trajectory of monitored quantum walks without considering a noise model. Second, due to finite resolution effects (finite number of measurements) new metastable quantization effects are found.
These effects, while very visible in our experiments, are expected to vanish in the limit of infinite number of measurements, giving new insights into monitored quantum systems.

\begin{figure}
\begin{center}
\includegraphics[width=0.8\linewidth]{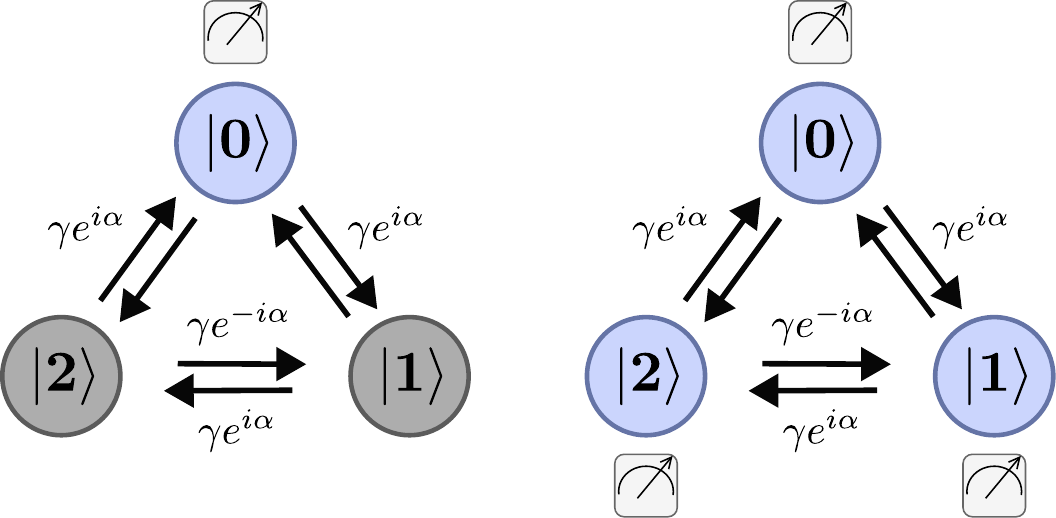}
\end{center}
\caption{
Scheme of the tight-binding model for a ring with three sites ($\ket{\bf 0}$, $\ket{\bf 1}$ and $\ket{\bf 2}$) pierced by a
magnetic flux  $\alpha$, corresponding to a directed or chiral triangle graph with complex edge weights, for two different measurement protocols. $\gamma$ denotes the strength of the hopping matrix element.
Left panel:
The on-site protocol measures periodically only the target state $\ket{\bf{0}}$. Right panel: The tracking protocol measures periodically all sites. The measurement is indicated with a measuring device.
In both cases, the hitting time is the first time
when the system is detected in state $\ket{\bf{0}}$.}
\label{fig:ring}
\end{figure}

The paper is organized as follows. Sec.~\ref{Sec:Meas} is the theoretical part that describes the model, the measurement protocol, and how to implement it on a quantum computer. 
In Sec.~\ref{Sec:Theory} and Sec.~\ref{Sec:Exp} we present the asymptotic theory and experiments, respectively, for the hitting time when the initial and target states are identical for different control parameters.
 In Sec.~\ref{Sec:dark} we vary the initial state, which leads to interference effects and dark states.
 Finite resolution effects are discussed in Sec.\ref{Sec:disc}.
We summarize our results in Sec.~\ref{Con} and propose some ideas for future studies.

\section{Model and Measurement Protocols \label{Sec:Meas}}
\subsection{Model}
We consider a tight-binding chiral quantum walk \cite{chiral} on a triangle graph with complex edge weights in the presence of a magnetic
flux (Fig.~\ref{fig:ring}) for two different monitoring protocols. 
 The localized states
are $\ket{\bf{0}},\ket{\bf{1}}$ and $\ket{\bf{2}}$ 
and the initial state is $\ket{\psi_{{\rm in}}}$. 
 The measurement-free evolution
is defined via unitary dynamics
\begin{eqnarray}
 U =\exp(- i H \tau) 
\end{eqnarray}
 with the time intervall $\tau$ between detection attempts, and $\hbar=1$. The Hamiltonian $H$
is
\begin{equation}
\label{eq1}
H = -\gamma \ e^{i \alpha } \left( \ket{\bf{0}} \bra{\bf{1}} + \ket{\bf{1}} \bra{\bf{2}}+\ket{\bf{2}} \bra{\bf{0}} \right) + {\rm h.c.} \ .
\end{equation}
Here h.c. stands for hermitian conjugate. $\alpha$ and $\gamma$ are control parameters that model the effect
of magnetic flux and hopping amplitude, respectively.
The time-reversal symmetry is broken when $\exp(i \alpha) \neq 1$.
The eigenvalues of the Hamiltonian are
\begin{equation}
\label{eigenvalues1}
E_k= - 2 \gamma \cos\left( {2 \pi k \over 3} + 
\alpha\right),  \ k=0,1,2 \ .
\end{equation}
When $\alpha=0$, we have a pair of distinct energy levels, and 
more generally when the flux $\alpha\neq 0$, we typically remove the two-fold degeneracy and the number of distinct energy levels is three \cite{PhysRevX.13.021021}.
\begin{figure}\begin{center}
\includegraphics[width=0.8\linewidth]{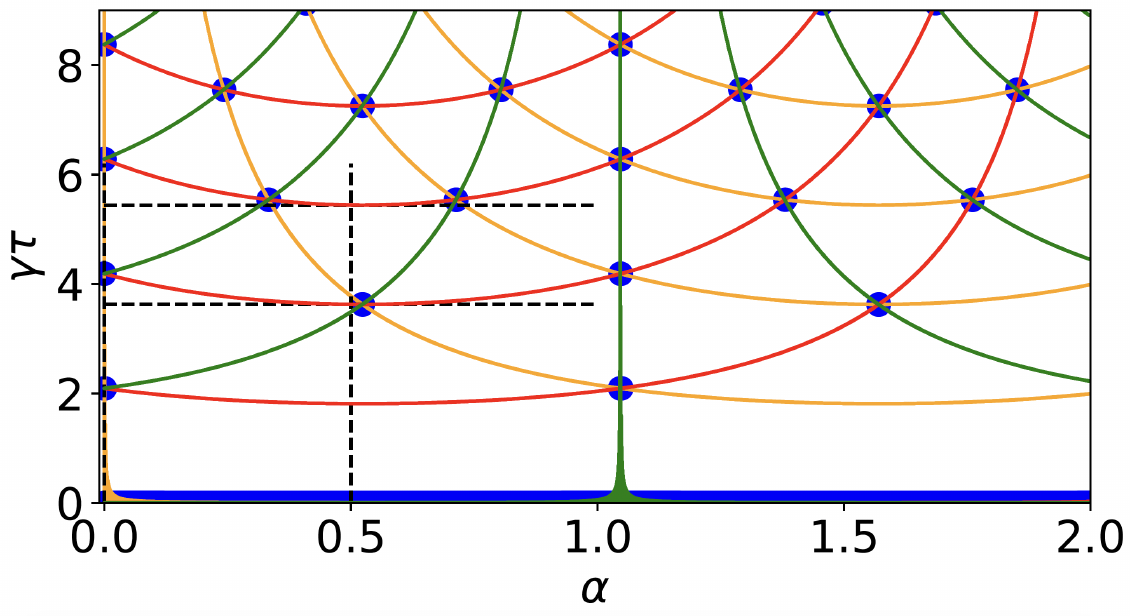}
\end{center}
\caption{
Parameters where two or three eigenvalues of the unitary $U$ are degenerate (phase factor matching diagram). In the plain $(\gamma \tau,\alpha)$, the phase factors
$\exp(-i E_k \tau)$ with $k=0,1,2$  match in pairs or triplets.
 Colors describe the
matching of two phase factors, for example, $\exp(-i E_0 \tau)=\exp(-i E_1 \tau)$
in red, and similarly for the pairs $(E_0,E_2)$ (green) and $(E_1,E_2)$ (orange). The matching of all three phase factors is shown as blue circles or a blue line at $\gamma \tau = 0$ indicating the Zeno regime.   
 Examples where three phase factors match are $(\alpha, \gamma \tau) = (0, 0), (0, 2 k \pi/3), (\pi/6, 2 \pi/\sqrt{3} ), (\pi/3, 2 k \pi/3), (\pi/2, 2 \pi/\sqrt{3} ), ....$. 
 The two dashed vertical lines ($\alpha = 0$ and $\alpha = 0.5$) as well as the horizontal bands ($\gamma \tau = 2 \pi/\sqrt{3}$ and $\gamma \tau = 3 \pi/\sqrt{3}$) indicate the parameter regime considered for the computation of the first hitting return time on IBM {\it Sherbrooke}.
 The phase factor matching diagram is obtained experimentally by studying the dark states (on-site protocol) in Fig.~\ref{fig3D_Ex}~(c).}
\label{fig:phases}
\end{figure}
The removal of energy degeneracy will have a profound effect on the mean return time and the dark states in the system as we will soon show.

To understand the physical properties of the monitored quantum walks, 
it is useful to visualize at which parameters $\alpha$ and $\gamma \tau$ (our control parameters), 
the eigenvalues of $U$ are degenerate
\begin{eqnarray}
 \exp (-i E_k \tau) = \exp(-i E_l \tau) \ ,  (k \neq l), \ k,l = 0,1,2   
\end{eqnarray}
in a phase factor matching diagram; see Fig.~\ref{fig:phases}. The analytical results are given in Appendix~\ref{phase}. For $\alpha = 0$ two phase factors match due to the degeneracies of the eigenenergies, while three phase factors match for $\gamma \tau =2\pi j/3  $ with $j \in \mathbb {N}$.
Other points where three phase factors match are, e.g., $(\alpha, \gamma \tau) = (\pi/6, 2\pi/\sqrt{3}), (\pi/3, 2 \pi j/3 ), (\pi/2,2\pi/\sqrt{3})$, etc.
 We will later see how this matching phase factor diagram is related to the hitting time problem as well as to the dark states.
We define a target state $\ket{\bf{0}}$ which we monitor
each $\tau$ units of time using two measurement protocols. These are described in the next two subsections.

\subsection{On-site protocol}

First, we consider the stroboscopic
measurement scheme \cite{Friedman2017}, which we call the local measurement protocol or the on-site
protocol.
Here, measurements are performed every $\tau$ units of time. Between measurements the dynamics is controlled by the unitary time evolution $U =\exp(- i H \tau) $.
The measurement is made locally at $\ket{\bf{0}}$ to detect this state; see Fig.~\ref{fig:ring} (left panel).
Here, the measurements yield either a no
or a yes, that is, the quantum walker is either not found or is found on $\ket{\bf{0}}$.  Mathematically,
the measurement is described by the projector $\ket{\bf 0} \bra{\bf 0}$.
A typical string of measurements is, for instance,
$$ \{ \mbox{no},
\mbox{no}, \mbox{no}, \mbox{yes}, \cdots \}.$$
In this case, the first hitting
time is $4 \tau$. The first click yes gives the random
number of measurements, denoted $n$,  needed to detect the quantum walker in the
target state $\ket{\bf{0}}$. In this case, unless we deal with a two-state system,
the information obtained in this way does not specify the state function
of the walker after each null measurement. Hence, after obtaining a no-result, we do not know what the amplitudes are for states $\ket{\bf{1}}$ 
and $\ket{\bf{2}}$.

\subsection{Tracking protocol}

 Second, we consider the tracking protocol \cite{Didi},  with the  target state $\ket{\bf{0}}$ where we record every
$\tau$ units of time the position of the walker on the graph; see Fig.~\ref{fig:ring} (right panel).
In this stroboscopic protocol, measurements are made at times
$ \tau, 2 \tau, \dots, 20 \tau$.
 In a typical realization of the process, we find a 
result, which is given by the eigenvalues of the position operator,  for example
$$ \{\bf{1},\bf{2},\bf{0},\cdots\}$$
or using the same initial condition and the same unitary
$$ \{ \bf{1},\bf{2}, \bf{1}, \bf{2}, \bf{1}, \bf{0},\cdots\}.$$
The first hitting time is clearly random and in these two examples it
is $3 \tau$ and $6 \tau$,
respectively. 

From the basic postulates of quantum measurement theory \cite{cohen1977quantum}, 
we know the state of the system immediately after each measurement.
This will be the eigenstate corresponding to the eigenvalue just recorded, assuming strong measurements.
For the first example, we will therefore find $$ \{\ket{\bf{1}},\ket{\bf{2}},\ket{\bf{0}},\cdots \}.$$
This implies that with this type of measurement we know precisely what the state of the system along the measured path is.
\subsection{The return problem}

In the return problem the target state and the initial state are the same.
This choice is also called the reccurence time problem. It will be studied with both protocols. 
Thus, we start with state $\ket{\psi_{\rm in}} =\ket{\bf{0}}$. 
Just before the first measurement at time $\tau$, the state function
is $\ket{\psi(\tau)}= U \ket{ \psi_{\rm in} }$. For the tracking protocol, we then measure the location
of the walker on the triangle graph. Suppose that the first measurement yields $\bf{1}$,
i.e., the system is projected to state $\ket{\bf{1}}$.
We then continue and
evolve the system until the second measurement according to $U \ket{\bf{1}}$,
and measure again. For the on-site protocol the new state is the projection to the other accessible states if we do not detect the quantum walker on site $\ket{\bf 0}$.
Note that when we study dark state physics, we use other initial conditions,
see below.

\subsection{Implementation on the Quantum Computer}

To record the trajectory on a quantum computer we have to encode the tight-binding Hamiltonian into a qubit Hamiltonian:
\begin{eqnarray*}
    H = -\frac{1}{2}\gamma [\cos(\alpha)(\sigma_{1}^{x}+\sigma_{2}^{x}+\sigma_{1}^{z} \sigma_{2}^{x}+\sigma_{1}^{x} \sigma_{2}^{z}+\sigma_{1}^{x} \sigma_{2}^{x}+\sigma_{1}^{y}  \sigma_{2}^{y}) \nonumber\\
    +\sin(\alpha)(\sigma_{1}^{y}-\sigma_{2}^{y}+\sigma_{1}^{y} \sigma_{2}^{z}-\sigma_{1}^{z} \sigma_{2}^{y}+\sigma_{1}^{x}  \sigma_{2}^{y}-\sigma_{1}^{y} \sigma_{2}^{x})]
\end{eqnarray*}
where $\sigma_x$, $\sigma_y$ and $\sigma_z$ are the Pauli matrices. The Hamiltonian $H$ defines two disconnected subspaces, the first composed of states $\ket{00},\ket{01},\ket{10}$
and the second of $\ket{11}$. 
The unitary evolution operator $U(\tau) = \exp(-i H \tau)$ is decomposed into unitary gates using Cartan's decomposition \cite{PhysRevA.69.010301}.
This allows us to vary $\tau$  without increasing the depth of the circuits. 

For the on-site protocol, 
we need to detect the state $\ket{\bf{0}}$ while we do not want to receive the information that allows us to distinguish the states $\ket{\bf{1}}$ and $\ket{\bf{2}}$.
This can be achieved by measuring only one qubit, which we define as the right qubit in the Dirac notation. We use the following mapping between the qubits and the representation of the spatial states:
$\ket{01} \rightarrow \ket{\bf{0}}, \ket{10} \rightarrow \ket{\bf{2}} \ \mbox{and} \
\ket{00} \to \ket{\bf{1}}$. Measuring the right qubit in state $\ket{0}$ does not give any information to distinguish the states $\ket{\bf{1}} = \ket{10}$ and $\ket{\bf{2}} = \ket{00}$ but measuring the right qubit in state $\ket{1}$ the system is in $\ket{\bf{0}} =\ket{01} $ with certainty. 

For the tracking protocol, 
we measure the states of both qubits, and thus after each measurement we
find the outcome $00$, $01$ or $10$.
For error prevention reasons, we use a different mapping between the qubit and the spatial state representation:
$\ket{00} \rightarrow \ket{\bf{0}}, \ket{01} \rightarrow \ket{\bf{1}} \ \mbox{and} \
\ket{10} \to \ket{\bf{2}}$.
The reason is that experimentally there is an asymmetry in the measured spectrum. For example, it is much more likely to migrate from a state $\ket{01}$ to $\ket{00}$ than from $\ket{00}$ to $\ket{01}$ due to decay processes. Therefore, we define $\ket{00}$ as our initial state.

As an error suppression strategy, we use dynamical decoupling by inserting two $X$-gates at specific intervals on the qubit, which is not measured, to keep it coherent \cite{PhysRevApplied.20.064027}. 
 We compute the time for the first detection or the first hitting time $n \tau$ and evaluate its statistical properties as a function of $\alpha$ and $\gamma$.
Effectively, the measurement
process has ended
when we find the target state for the first time. 
The result obtained from the measurements is a string of size $N$ 
which yields what we call a trajectory (a
string of $N$ bits for the on-site protocol and $N$ times two bits for the tracking protocol for 32,000 runs). 
In practice, we cannot
stop the experiments in the middle of the process, i.e.,
currently there is no conditional abort option on IBM quantum computers. We therefore shorten the trajectories using classical post-processing at the bit position where the target site was measured. 
\subsection{Observables}
Now we define statistical observables of interest. They can be estimated
from the sample paths and will depend on the parameters of the model, such as $\gamma \tau$, $\alpha$, the initial state and the measurement protocol.
  Let $F_n$ be the probability that the quantum walker is detected for the first time
in the target state at the $n$-th measurement \cite{Friedman2017,Didi}. Thus $F_1$ is the probability
that the target state 
is recorded at the first measurement event, while
$F_2$ describes the case where the state was recorded for the first time in the second attempt.
$\langle n \rangle$ is the conditional mean number of events until detection
\begin{eqnarray}
    \langle n (N)\rangle  = \frac{\sum_{n=1}^N n F_n}{\sum_{n=1}^N F_n} \ ,
    \label{eq:condn}
    \label{q:mean}
\end{eqnarray}
for $\sum_{n=1}^N F_n \neq 0$. Hence, $\tau \langle n (N) \rangle$ is the mean time until detection.
From now on, we will call this physical quantity mean return time for brevity.
We also define the total detection probability $P_{{\rm det}}$
\begin{eqnarray}
    P_{{\rm det}}(N) = \sum_{n=1}^N F_n \ ,
    \label{eq:pdet}
\end{eqnarray}
which is the probability that a detection
event was recorded. The estimation of this probability comes from the number of strings without detection.
In an ideal situation with infinite resolution,  these are obtained through
$\lim_{N \to \infty} P_{{\rm det}}(N) = \sum_{n=1} ^\infty F_n$ and if the latter is unity, that is, if the system is detected with probability one, 
$\lim_{N \to \infty}  \langle n (N)\rangle  = \sum_{n=1} ^\infty n F_n$. 
As mentioned in the Introduction, even for the triangle model under study we
can find cases where $P_{{\rm det}}< 1$  due to destructive interference,
as explained below.

In the experiment, we obtain the statistical properties of the first
detection time for a finite number of measurements $N$ giving rise to finite resolution effects. Per trajectory, the first
detection time is $n \tau$, where the integer $n$ is the random number of measurements until the first detection of the target state.
Clearly in our experiments $n \le N$.
  Note that the obtained strings are vectors of size $N$.
It is possible that the string did not contain the target state $\ket{\bf{0}}$, a fact that will become important later, close to topological transitions.
 
\section{Theory recap \label{Sec:Theory}}

 We consider the return problem when the initial state $\ket{\psi_{{\rm in}} } = \ket{ \bf{0}}$ is detected for the first time. 
Furthermore, for now we assume
$N \to \infty$.
\subsection{On-site protocol (theory)}

The theorem of recurrence for on-site measurements reads \cite{Gruenbaum2013,Grunbaum2014} 
\begin{equation}\label{meanen}
\langle n \rangle = \mbox{number of distinct phase factors} \ \  \exp(- i E_k \tau)
\end{equation}
where $E_k$ are the eigenstates of $H$ [Eq.~(\ref{eigenvalues1})] for the triangle model. 
From Eq.~(\ref{eigenvalues1}) it becomes clear why the phase factor matching diagram is a very useful tool. For example, for the
choice of parameters corresponding to blue circles in Fig.~\ref{fig:phases} we expect $\langle n \rangle = 1$,
while for the colored lines (besides blue) $\langle n \rangle = 2$, and anywhere else $\langle n \rangle = 3$.
This holds in general
under the condition that the energy states, i.e., eigenstates of $H$,
denoted $\ket{E_k}$ have finite overlap with the detected state. 
For example, in the Zeno limit $\tau \to 0$, the three phase factors merge, and hence $\langle n \rangle$
 is equal to one. This is expected for the return problem, since we start at the measured state, and hence the first
measurement detects it when $\tau =0$. 
Phase factor matching conditions correspond to removal of a state from the effective Hilbert space. If phase factor match
$\exp(- i E_1 \tau) =\exp(-i E_2 \tau)$, the state
$\psi \sim \langle {\bf 0}|E_1 \rangle |E_2 \rangle - \langle {\bf 0}|E_2 \rangle |E_1 \rangle$
cannot be detected. Thus, the phase factor matching choices of $\gamma \tau$ and $\alpha$ reduce the dimension of the Hilbert space, which can be shown to lead to faster detection compared to the cases where the phase factors do not match.

We provide more details in Appendix~\ref{appendix:onsite}.

\subsection{Tracking protocol (theory)}

Except for special model parameters, the mean return time $\langle n\rangle$ is the number of states
in the system, which is three for the triangle model under study. The result $\langle n\rangle=3$
can be viewed
as a classical result. More specifically,
consider a classical random walk coupled to a heat bath in the infinite
temperature limit. In this case, in a steady state,
all states are equally likely, namely the occupation probability is $1/3$.
Then from the classical Kac theorem for the return problem $\langle n \rangle = 3$ \cite{Kac}. 
At least in the classical domain, this is easy to understand, since the probability of measuring the
walker in the target state in the first measurement is $1/3$, in the second
$(2/3)(1/3)$ etc. and hence $\langle n \rangle =\sum_{n=1} ^\infty n (2/3)^{n-1} (1/3)=
3$.  Thus, tracking, for most of the choices of $\tau$,
drives the system to a classical limit.  
The quantum aspect of the hitting-time process
is found for special values of $\tau$ that capture the revival of the wave packet and hence the underlying periodicity of the quantum dynamics.
Namely, there exist revival times where the initial wave function returns
to its original state in addition to an unimportant phase. Then if we measure
at that time, the quantum walker is detected with probability one in the target state in the first
measurement and hence $\langle n \rangle =1$.
In particular, when $\tau=0$ we have $\langle n \rangle =1$. In the diagram in Fig.~\ref{fig:phases} these special sampling times correspond to three phase factors matching, shown as blue dots and blue lines. 
Clearly, the idealized theory predicts that as we vary $\gamma \tau$ and $\alpha$, when three phase factors match, we will see a dip or a transition in $\langle n \rangle$. Thus, $\langle n \rangle$ will jump from $\langle n \rangle = 3$ to $\langle n \rangle = 1$ and back. 
In Appendix~\ref{appendix:Didi}, 
following Ref.~\cite{Didi} we provide more details of the theory by introducing the corresponding Markov or stochastic matrix.

\section{First hitting return times on IBM quantum computers\label{Sec:Exp}
}

We now turn to the experiments for the first hitting return time where for the initial condition we start at the target state $$\ket{\psi}_0 = \ket{\bf{0}}.$$
Theoretically, $\langle n \rangle$ exhibits pointwise discontinuous behavior as mentioned.
Can we see this
on a quantum computer? At first glance, this might seem hard
since the width of these transitions is theoretically zero, but in reality for finite $N$ the transitions are broadened. In the following we compare the experimental values, the classical simulation of Eq.~\ref{eq:condn} and the asymptotic theory ($N \rightarrow \infty$).

\begin{figure}\begin{center}

\includegraphics[width=1\linewidth]{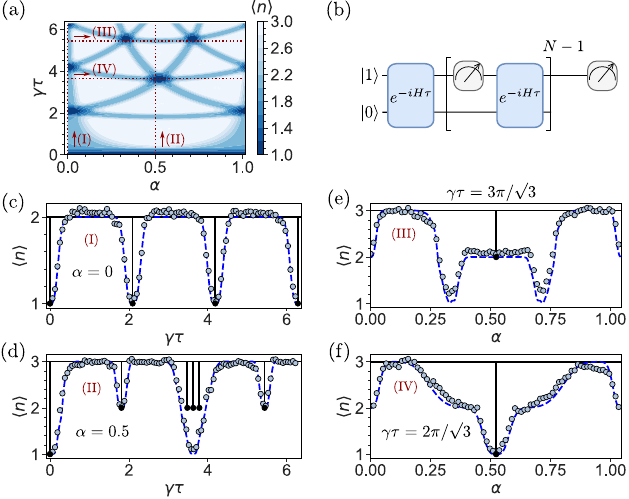}

\end{center}
\caption{
(a) Classical simulation of the mean return time $\langle n \rangle$ for $N = 20$ for the on-site protocol 
as a function of $\gamma \tau$ and $\alpha$. For $\langle n \rangle = 1$ three phase factors match (dark blue), for $\langle n \rangle = 2$ two phase factors match (medium blue) and for $\langle n \rangle = 3$ all phase factors are different (white).
The paths through the parameter space labeled as (I), (II), (III) and (IV) are indicated with red dashed lines and correspond to (c), (d), (e) and (f), respectively. Compared to the phase factor matching diagram in Fig.~\ref{fig:phases} the lines are broadened due to finite $N$. 
(b) Quantum circuit for two qubits representing the three localized states for the on-site protocol with the initial state $ \ket{\bf{0}} = \ket{01}$, where the unitary $U=\exp{(-iH \tau)}$ and measurements are applied alternately. Only the upper qubit is measured, since
we need to detect the state $\ket{\bf{0}}$ while we do not want to receive the information that allows us to distinguish the states $\ket{\bf{1}}$ and $\ket{\bf{2}}$.
(c) - (f) Mean return time $\langle n \rangle$ 
 recorded on IBM {\it Sherbrooke} (light blue circles), obtained with the asymptotic theory for $N \rightarrow \infty$ (black solid lines) and simulated for $N = 20$ (blue dashed lines): 
  (c) [path (I)] $\langle n \rangle$ versus $\gamma \tau$
 for $\alpha=0$.
(d) [path (II)] When $\alpha = 0.5$ we almost always find $\langle n \rangle =3$.
As explained in the text, this is related to the removal of energy-level degeneracy
when the magnetic flux is turned on. Note that
some fine structure details predicted by the asymptotic theory, for $\alpha = 0.5$, are washed out in the experiment, see the three nearby dips that merge into one resonance here. 
(e) [path (III)] Mean return
time $\langle n \rangle $ as a function of $\alpha$ for $\gamma \tau = 3\pi/\sqrt{3}$. $\langle n \rangle$ develops a plateau, as well as additional transitions that are absent for $N \rightarrow \infty$. (f) [path (IV)] Mean return
time $\langle n \rangle $ for $\gamma \tau = 2\pi/\sqrt{3}$ as a function of $\alpha$. In this example clearly the asymptotic theory is not predictive, while finite time simulations agree with experiments.  }
\label{fig:exponsite}
\end{figure}
\begin{figure}\begin{center}

\includegraphics[width=1\linewidth]{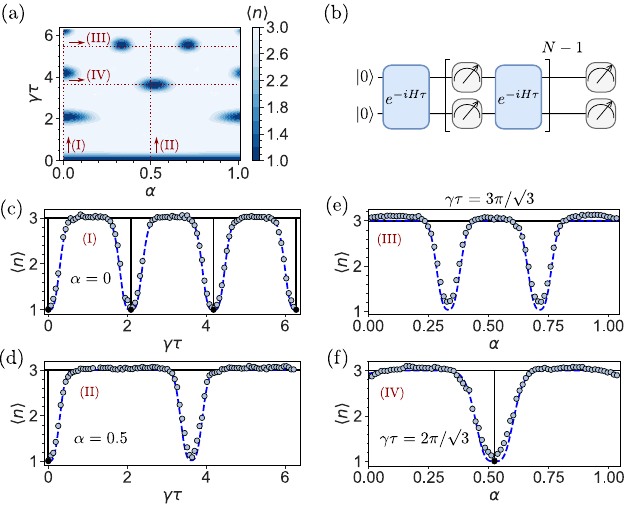}

\end{center}
\caption{
(a) Classical simulation of the mean return time $\langle n \rangle$ for $N = 20$ for the tracking protocol 
as a function of $\gamma \tau$ and $\alpha$. 
For $\langle n \rangle = 1$ three phase factors match (dark blue), and for $\langle n \rangle = 3$ all phase factors are different (white).
Paths through the parameter space labeled (I), (II), (III) and (IV) are indicated with red dashed lines and correspond to (c), (d), (e), and (f), respectively. Compared to the phase factor matching diagram in Fig.~\ref{fig:phases} the dark blue areas are broadened due to finite $N$.   
(b) Quantum circuit for two qubits representing the three localized states for the tracking protocol with the initial state $ \ket{\bf{0}} = \ket{00}$, where the unitary $U=\exp{(-iH \tau)}$ and the measurements are applied alternately. 
(c) - (f) Mean return time $\langle n \rangle$ 
 recorded on IBM {\it Sherbrooke} (light blue circles), obtained with the asymptotic theory for $N \rightarrow \infty$ (black solid line) and simulated for $N = 20$ (blue dashed line): 
  (c) [path (I)]  $\langle n \rangle$ versus $\gamma \tau$
 for $\alpha=0$.
$\langle n \rangle$
is quantized ($\langle n \rangle = 3$) and its value, for almost any choice of $\gamma \tau$,  
differs by unity from
the on-site measurement protocol, the case presented in Fig.~\ref{fig:exponsite} (c) when $\langle n \rangle = 2$.
(d) [path (II)] 
For finite magnetic flux $\alpha = 0.5$ the transitions to $\langle n \rangle =1$ are absent except in the Zeno limit at $\gamma \tau = 0$ for $N \rightarrow \infty$. This is different for $N =20$.
The result for $\alpha=0.5$ shows a transition to approximately $\langle n \rangle =1$ at $\gamma \tau = 3.63$. 
(e) [path (III)] Mean return
time $\langle n \rangle $ for $\gamma \tau = 3\pi/\sqrt{3}$. For $N = 20$ $\langle n \rangle$ develops additional transitions that are absent for $N \rightarrow \infty$ (f) [path (IV)] Mean return
time $\langle n \rangle $ for $\gamma \tau = 2\pi/\sqrt{3}$ as a function of $\alpha$. 
We see a clear broadening effect of the resonance,  while in (d) and (e) broadened resonances show up, which are not present in the asymptotic theory. }
\label{fig:exptracking}
\end{figure}

\subsection{On-site protocol (experiment)}
Fig.~\ref{fig:exponsite}~(a) shows the parameter regime (red dashed line) utilized for the computation of the mean return time as well as the classical
 simulation of the mean return time as a function of $\gamma \tau$ and $\alpha$ for $N = 20$. In comparison to Fig.~\ref{fig:phases} the areas for $\langle n \rangle = 1$ (dark blue) and $\langle n \rangle = 2$ (medium blue) are broadened. 
The figure clearly shows how the conditions for phase factor matching can lead to jumps in $\langle n \rangle$.

 For the experimental computation of $\langle n \rangle$ we utilize the quantum circuit in Fig.~\ref{fig:exponsite}~(b). The on-site  protocol can be emulated by measuring the upper qubit after each unitary $U$. 

Let us first discuss the results of the asymptotic theory using Eq.~(\ref{meanen}).
In Fig.~\ref{fig:exponsite} (c) - (f), 
we consider $\langle n \rangle$ versus the dimensionless
sampling time $\gamma \tau$, as well as the magnetic flux $\alpha$ (black solid line).
For almost any $\tau$, on average, we return after two measurements, i.e., $\langle n \rangle= 2$. This holds
when the magnetic flux is turned off ($\alpha=0$). 
In contrast,
 when $\alpha \neq 0$ we have besides very special choices of $\tau$,
 $\langle n \rangle = 3$ which is related to the fact that the magnetic flux  $\alpha>0$ removes the degeneracy by breaking the time-reversal symmetry and the rotational invariance. 
In addition, one observes typical sudden plunges in $\langle n \rangle$. These
are found for special values of $\gamma \tau$ and $\alpha$. As mentioned in the Introduction, these transitions, e.g., dips in $\langle n \rangle$ when plotted versus $\gamma \tau$ or $\alpha$
are related to a discontinuous change of the topology of the underlying
generating
function \cite{Gruenbaum2013,Grunbaum2014,Yin2019}. 

Quantization of the mean return time of the experimental result (light blue circles) is clearly evident for most values of
$\gamma \tau$ and $\alpha$. Far from transitions in $\langle n \rangle$, theory
predicts, and the experiments confirm, that for on-site measurements, with and
without magnetic flux,
$\langle n \rangle \approx 2$ [Fig.~\ref{fig:exponsite}~(c)] and $\langle n \rangle \approx 3$ [Fig.~\ref{fig:exponsite} (d)-(f)], respectively. Notwithstanding these successes, theory ($N \rightarrow \infty$) and experiment depart close to the parameters in which the broadening of the dark blue and medium blue areas is obvious in Fig.~\ref{fig:exponsite}~(a). 
When we choose path (III), in the ($\alpha$, $\gamma \tau$) plain, we are in the vicinity of a phase factor matching curve, though strictly not exactly on it. Since the number of measurements $N$ is finite in the experimental study, we are effectively witnessing a topological state with $\langle n \rangle = 2$, instead of $\langle n \rangle = 3$ predicted by theory, see Fig.~\ref{fig:exponsite}~(e). 
We observe not only a broadening of the resonances, but also additional resonances for finite $N$ that are not present in the theoretical graph ($N \rightarrow \infty$) [Fig.~\ref{fig:exponsite} (d)].

Furthermore, experiment and asymptotic theory do not match when 
triplets of dips of $\langle n \rangle$ are in the vicinity of one another on path (II).  
Indeed, when $\alpha=0.5$, we see from the theoretical prediction in Fig.~\ref{fig:exponsite}~(d) three nearby dips (black solid line), while in reality, we see one wide resonance (light blue circles). Thus, fine details are wiped out near the point where all three phase factors are almost matching.

What is the cause of the deviation between theory
(black solid lines) and experiment (light blue circles)
close to these topological transitions?
It might be due to noise, readout errors, or finite resolution
effects. Without a doubt, all of these effects may play some
role.
However, for the mean return time under study, we concluded that the main factor is the finite resolution
of the experiment.
The unitary dynamics and measurements
implemented on the quantum computer
work well on the time and size scale of our experiment, indicated by the fact that the experimental points lie close to the simulation on a classical computer (blue dashed lines). Finite resolution effects are discussed in detail in Sec.~\ref{Sec:disc}.

\subsection{Tracking protocol (experiment)}

Fig.~\ref{fig:exptracking}~(a) shows the parameter regime (red dashed line) utilized for the computation of the mean return time as well as the
 simulation of the mean return time as a function of $\gamma \tau$ and $\alpha$ for $N = 20$ for the tracking protocol. Compared to Fig.~\ref{fig:phases} the areas for $\langle n \rangle = 1$ (dark blue) are broadened and in comparison to the on site protocol Fig.~\ref{fig:exponsite}~(a) we observe the
dark color ($\langle n \rangle$) more sporadically, i.e., the lines where only two phase factors match are irrelevant.
 For the experimental computation of $\langle n \rangle$ we implement the quantum circuit in Fig.~\ref{fig:exptracking}~(b), where both qubits are measured after applying the unitary $U$.

We start by discussing the theory for $N \rightarrow \infty$ (black solid line).
The tracking protocol gives with and
without flux ($\alpha$) $\langle n \rangle = 3$, for nearly all $\tau$, according to theoretical predictions. 
Note that any small deviation of $\tau$ from the special revival times or $\tau=0$, that is, the Zeno limit, yields $\langle n \rangle = 3$ 
as shown in Fig.~\ref{fig:exptracking} (c) - (f) as a black solid line.
  For the finite magnetic flux $\alpha = 0.5$ in Fig.~\ref{fig:exptracking} (d), there are no special sampling times $\tau$ except $\tau = 0$. (Similar for $\gamma \tau = 3 \pi/\sqrt{3}$ and finite $\alpha$ in Fig.~\ref{fig:exptracking} (e)). 
When three phase factors merge, the deviations between the theory valid for $N \to \infty$
and experiments (light blue circles) for $N = 20$ are large. 
We obtain a large deviation for the parameters in which the broadening of the dark blue areas is observed in Fig.~\ref{fig:exptracking}~(a) between experiment and assymptotic theory. The broadening is present in direction of $\gamma \tau$ and $\alpha$ which leads not only to a broadening of the resonances, but also to additional resonances for finite $N$ that are not present in the theoretical graph ($N \rightarrow \infty$) [Fig.~\ref{fig:exptracking} (d) and (e)] but in the finite $N$ simulation (blue dashed line). 
Similarly to the on-site protocol, the deviation of the experimental values from the asymptotic theory is due to finite-resolution effects. These are discussed in detail in Sec.~\ref{Sec:disc}.

\section{Dark States on IBM quantum computers \label{Sec:dark}}

 So far we have considered the return problem, where the initial state was
also the target state.
 As mentioned in the Introduction,
considering more general initial conditions, we encounter dark states,
and the eventual detection of the walker, even after an infinite number of repeated measurements, is not generally guaranteed \cite{Todd2006,Thiel2020}.

The detection probability is defined in Eq.~\ref{eq:pdet}.
As mentioned in the Introduction
for classical random walks on finite graphs like the triangle  model
$\lim_{N \to \infty} P_{{\rm det}}(N)=1$,
namely, the walker is detected with probability one.
In our study, considering a finite graph, whenever $\lim_{N\to \infty} P_{{\rm det}}(N)< 1$ we say that the system exhibits dark-state physics. Some initial conditions give $P_{{\rm det}}(N)=0$ and
these initial states are called dark states. 

\subsection{Dark states for zero magnetic flux}

Consider the case $\alpha=0$, 
the initial condition
\begin{equation}
\ket{\psi}_0= { \ket{\bf{1}} + e^{ i\phi} \ket{\bf{2}} \over \sqrt{2} } 
\label{eqIC}
\end{equation}
and as before the hitting process ends when we detect the system
in state $\ket{\bf{0}}$. We study the influence of the phase $\phi$ on the detection process.

For on-site measurement and when $\phi=0$ the system is detected with probability one, except for special values of sampling times. 
In contrast, if $\phi=\pi$,
the detection probability is zero. The latter case is caused by destructive interference and the symmetry of the problem. When $\phi=\pi$ the current from $\ket{\bf{1} }$ to the detected state is minus the current
from $\ket{\bf{2}}$, which means that for all times, including between measurements, the amplitude of the detected state is zero. 
This can break down when $\alpha \neq 0$, i.e., when the symmetry is broken. 
A different behavior is found
for the tracking protocol.
Then, 
after the first measurement, the wave packet is spatially localized in the system; hence, the symmetry is broken by the measurements, and the walker is eventually detected. 
Thus, a dark state in the on-site protocol can be bright for the tracking method.

\begin{figure}\begin{center}
\includegraphics[width=1\linewidth]{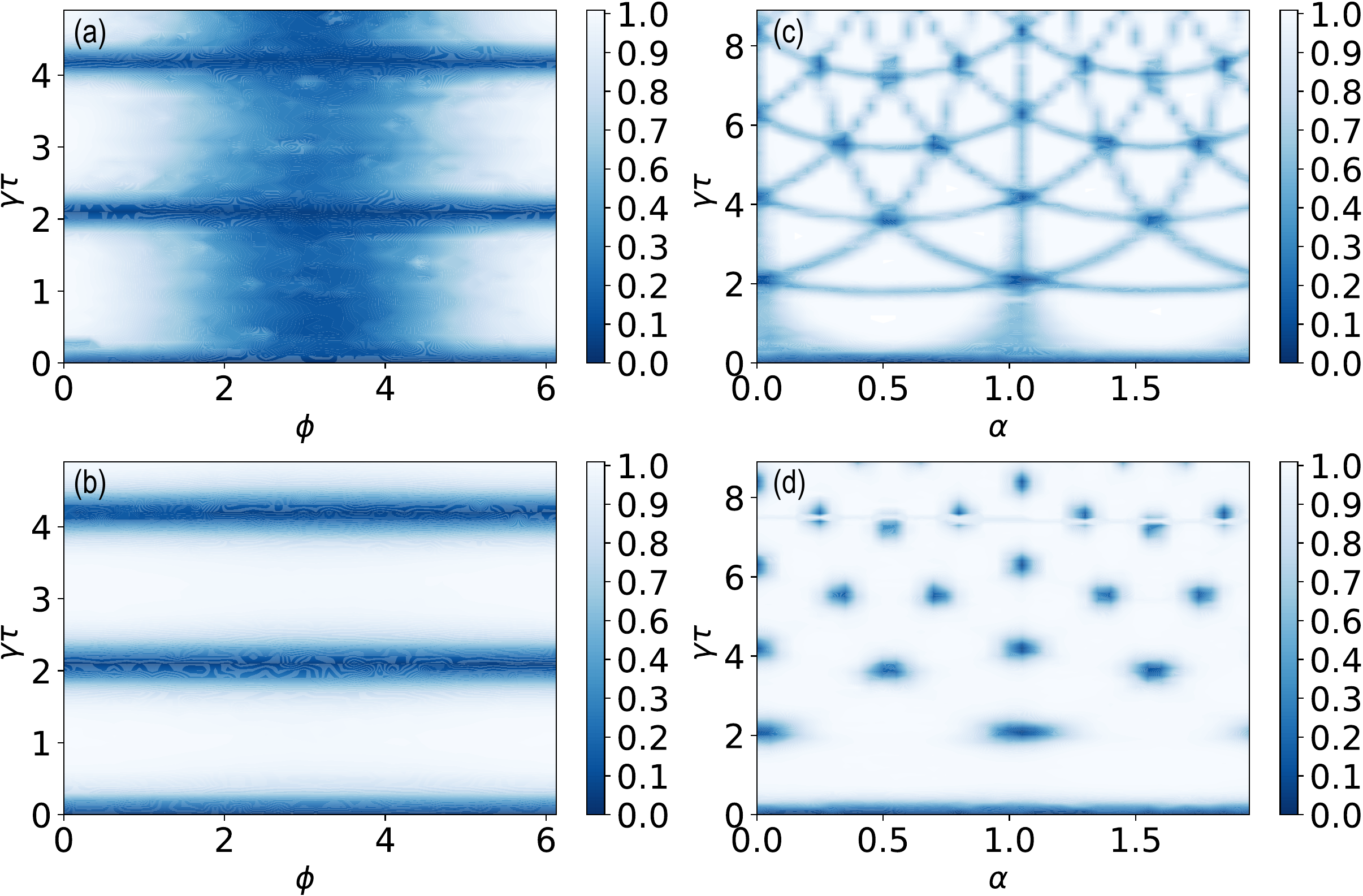}
\end{center}
\caption{ Detection probability
 $0 \le P_{{\rm det}}\le 1$ as a function
of the sampling time $\gamma \tau$ and the phase $\phi$ as well as $\gamma \tau$ and magnetic flux $\alpha$ obtained from IBM {\it Sherbrooke} with $N = 10$ for the on-site and tracking protocol. The dark blue color corresponds to $P_{\rm det} = 0$ (dark state), while the white color corresponds to $P_{\rm det} = 1$ (bright state).
 (a) $ P_{{\rm det}}$ as a function of $\gamma \tau$ and $\phi$ for the on-site protocol.
Dark states are found when $\phi = \pi$ is the result of destructive interference and for $\gamma \tau \to 0$ 
due to the Zeno effect.
Additional horizontal bands show dark states, which are found when the time of revival
in the initial
state is the same as the sampling time $\gamma \tau$ and $\alpha = 0$.
(b) $P_{\rm det}$ as a function of $\gamma \tau$ and $\phi$ ($\alpha = 0$) for the tracking protocol. 
The dark states found for $\phi=\pi$ for the on-site protocol turn bright
when the tracking method is used, 
since the latter breaks spatial symmetry,
while the former does not. The horizontal dark bands are due to revivals
that forbid detection of the system in the target state. (c) and (d) $P_{\rm det}$ as a function of $\gamma \tau$ and magnetic flux $\alpha$ ($\phi= 0$). The on-site protocol (c) and the tracking protocol
(d) exhibit vastly different trends, i.e.,
dark states are present in the tracking protocol where three phase factors match and in the on-site protocol where two and three phase factors match. As explained in the text, (c) yields the phase factor matching diagram shown in Fig.~\ref{fig:phases}, while (d) corresponds to three phase factors matching, which are indicated by blue circles in Fig.~\ref{fig:phases}.}
\label{fig3D_Ex}
\end{figure}

A comparison between the two protocols is presented in
the three-dimensional graphs in Fig. \ref{fig3D_Ex}~(a) and (b). 
The dark color in the figure corresponds to dark states. 
For the on-site protocol we find using methods in \cite{Thiel2020} 
\begin{equation}
\lim_{N \to \infty} P_{{\rm det}}(N) =
\left\{
\begin{array}{l l}
0, & \gamma \tau = 2 \pi k /3 \\
\ & \ \\
{1 + \cos(\phi) \over 2}, & \mbox{otherwise}
\end{array}
\right 
.
\label{eqPDET}
\end{equation}
The first line for $k=0$ corresponds to the well-studied Zeno limit. 
The Schr\"odinger equation being first order in time implies that the amplitude in the detected state $\ket{\bf{0}}$,
at time $\tau$, is proportional to $\tau$ but
according to the rules of quantum measurement theory, 
the probability of detecting the quantum walker is proportional to $\tau^2$, 
and hence the walker is not detected at all. 
It should be noted, however, that for any small value of $\tau$ and when $N \to \infty$ the walker is eventually detected, which means that the limits
of large $N$ and small $\tau$ do not commute. 
 Other values of $k=1,2,3,...$ are choices
of $\tau$ corresponding to revival times, when the wave function returns to its
original state, and hence the stroboscopic measurements cannot click yes even after many measurements and the quantum walker is never detected. 
The second line in Eq.
(\ref{eqPDET}) corresponds to the destructive interference that is strongest when $\phi=\pi$. 
It is a robust effect in the sense that, for on-site measurements,
it will hold for any choice
of measurement times. 

 We recorded $P_{{\rm det}}$ on a quantum computer
using on-site measurements; see
Fig.~\ref{fig3D_Ex} (a).
We use
$N=10$ measurements for each value of $\gamma \tau$ and $\phi$.
The three-dimensional plot, Fig.~\ref{fig3D_Ex}~(a),
clearly demonstrates the above mentioned features as horizontal and vertical dark stripes. In contrast,
in the tracking protocol [Fig.~\ref{fig3D_Ex}~(b)], we see only the horizontal lines corresponding to the Zeno effect and the revivals, since, as mentioned,
tracking breaks the symmetry in the system needed to maintain destructive
interference effects. 
 
Clearly, the results in Fig.~\ref{fig3D_Ex} (a) and (b) using the quantum computer and in Fig.~\ref{fig:3Dsim} in Appendix \ref{app:sim} 
using simulations indicate a very good agreement between
theory and experiment. 

\subsection{Dark states for a finite magnetic flux}

When $\alpha\neq 0$ the detection probability plots
presented so far,  could in principle be
replaced with a study  of $P_{{\rm det}}$
 versus of $\alpha,\phi$ and $\gamma \tau$.
To simplify the matter, in the experiments and simulations below, the initial condition
is $$\ket{\psi}_0 = \ket{\bf{1}}$$ and therefore the control parameters
are $\gamma \tau$ and $\alpha$.
In Fig.~\ref{fig3D_Ex}~(c) and (d), we plot the detection probability for the on-site and tracking protocols, respectively, and find a striking difference. 
For comparison, the corresponding simulation is shown in Fig.~\ref{fig:3Dsim} in Appendix \ref{app:sim}. 

 The main features of these figures can be explained as follows.
We start with the on-site protocol.
Consider the choice of parameters for which two phase factors merge. For example,
$\exp(-i E_1 \tau)= \exp(-i E_2 \tau)$. 
We also denote the corresponding eigenstates with $\ket{E_1}$ and $\ket{E_2}$, respectively.
Then an initial state 
\begin{equation}
\ket{\psi}_{{\rm in}} \sim \ \bra{\bf{0}} E_1 \rangle \ket{E_2} -
\bra{\bf{0}} E_2 \rangle \ket{E_1} 
\label{eqdark}
\end{equation}
is clearly orthogonal with respect to the detected state $\ket{\bf{0} }$.
In this case, every unitary 
step yields the same global phase shift $\exp(-i E_k \tau)$ with $k=1,2$.
Therefore, the amplitude of the target state will be zero for any subsequent measurement.
In other words, this initial condition
is completely dark and cannot be detected at all. 
Any initial state that is not orthogonal to this state cannot be detected with probability one; hence
in our example, the initial state under study is not bright when two phase factors merge. 

 When the three phase factors merge, a stronger effect will occur on the triangle graph. Then the 
wave function, 
 every $\tau$ units of time, returns to its initial state
and hence revives.
 Since in the case studied in this section, the initial state is orthogonal
to the detected state, clearly one cannot detect the walker.
The same holds for the tracking measurement. The walker cannot be detected
in the first measurement, nor in any of the repeated measurements. 
 Hence, for on-site measurements, nonbright states are found
when two or three phase factors merge.
For the tracking protocol, dark states are found only when three phase factors merge;
otherwise, the state is detected with probability one. 
 Therefore, to better understand the detection, we return to the
phase factor matching diagram in Fig.~\ref{fig:phases},
where we plot in the ($\gamma \tau$, $\alpha$) plane curves when
two phase factors coincide. 
That is,
$\exp(-i E_1 \tau) = \exp(-i E_2 \tau)$ gives one branch of this relation between
$\gamma \tau$ and $\alpha$
 and similarly for pairs $(E_1,E_3)$ and $(E_2,E_3)$. The plot of this information requires
the energy levels that depend, of course, on $\alpha$.
Different pairs of matching phase factors are presented in different colors.
 When three phase factors merge (circles), i.e., when curves cross each other in Fig.~\ref{fig:phases},
we see that for the tracking protocol, the detection
probability in Fig.~\ref{fig3D_Ex}~(d) is small or zero.
These crossing events appear as dark blue circles in Fig. \ref{fig3D_Ex}~(d).
 In the small $\gamma \tau$ limit, we see a clear dark stripe, 
representing Zeno dark states. When $\tau = 0$ the three phase factors coincide.

When two phase factors match,  
in Fig.~\ref{fig:phases}, we see for the on-site measurements, i.e., in Fig.~\ref{fig3D_Ex}~(c), the corresponding darker color. An issue for future study is whether one can distinguish between the colors in Fig. \ref{fig:phases}, using repeated measurements.
If we want to remove curves in Fig.~\ref{fig3D_Ex}~(c), at least partially,
then we have to choose an initial state orthogonal
to the dark-energy state in Eq.~(\ref{eqdark}) which will be bright corresponding to the merging of the two phase factors $\exp(-i E_1 \tau)=\exp(-i E_2 \tau)$. 
Adding a phase to the initial state, like in
Eq.
(\ref{eqIC})
 can also add features to these plots.
Can any of this be observed in the experiment? The results
presented in Fig.~\ref{fig3D_Ex} are very convincing in this regard. 

If we consider the return problem, the detection probability is
at its maximum, and
hence a figure like Fig.~\ref{fig3D_Ex}~(c) would be colored white and, therefore, not very informative. 
In the dark area where $P_{\rm det} = 0$ the mean return time is $\langle n \rangle = 1$ for both protocols. On brighter lines where the two phase factors match, we get $\langle n \rangle = 2$ (return) and $0 < P_{\rm det} < 1$ (transfer) for the on-site protocol. Also, here finite resolution effects are important, as for $N \rightarrow \infty$ the dark lines will become infinitely thin.

\begin{figure}
\begin{center}
\includegraphics[width=0.8\linewidth]{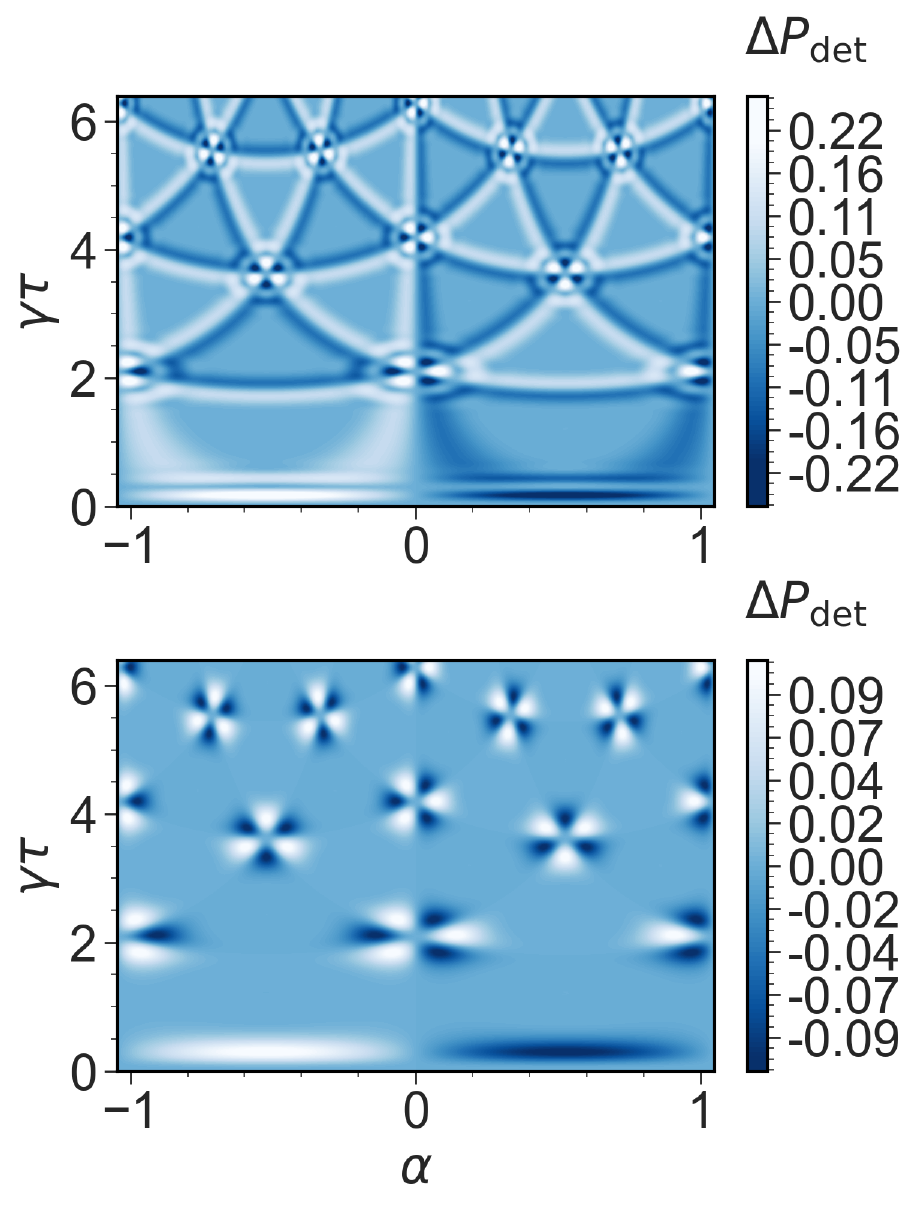}
\end{center}
\caption{Simulation of the difference of the total detection probability $\Delta P_{\rm det} = P_{\rm det}(\ket{\psi_{\rm in}} = \ket{\bf{1}} )- P_{\rm det}(\ket{\psi_{\rm in}} = \ket{\bf{2}} )$ with the different initial states $\ket{\bf{1}}$ and $\ket{\bf{2}}$, which are not the target states as a function
of the sampling time $\gamma \tau$ and magnetic flux $\alpha$ ($N = 10$) for the on-site (upper panel) and tracking protocol (lower panel).}
\label{fig:chiral}
\end{figure}

Finally, we pose the question of whether the total detection probability changes if we choose $\ket{\psi}_0 = \ket{\bf{2}}$. 
The total detection probability for the transition from $\ket{\bf{1}}$ to $\ket{\bf{0}}$ and $\ket{\bf{2}}$ to $\ket{\bf{0}}$, respectively, is different and depends on the direction of the magnetic flux $\alpha$ for $N = 10$.
  The difference of the total detection probabilities is shown in Fig.~\ref{fig:chiral}, where an asymmetry is observed around the broadened phase-matching areas and lines.
The total detection time of the chiral quantum walker for the transfer depends on the direction (clockwise or counter clockwise) and also on the sign of the magnetic flux.  
Breaking of the time-reversal symmetry can therefore enhance or suppress the total detection and therefore provides a directional control for certain parameters.

\section{Finite resolution \label{Sec:disc}}
Finite resolution plays an important role in explaining the deviations between finite-time experiments and the theory for $N \rightarrow \infty$. 
  It is present in the data of the mean return time $\langle n \rangle $ Fig.~\ref{fig:exponsite} and Fig.~\ref{fig:exptracking} as well as the dark states Fig.~\ref{fig3D_Ex}. 
By finite resolution, we mean that measurements are performed $N$ times per run. 
Typically, we record the target state for the first time in some random
measurement attempt numbered
$n_i \le N $.
We repeat this process by constructing an ensemble of trajectories
to sample the mean return time which is maximally the number of runs. On the quantum computer that we choose the number of runs equal to $32,000$.
We have already explained that in some cases we might not find the target within
these $N$ measurements. Hence, in the figures presented, we show sample averages that, in principle, depend on our choice of $N$.
More specifically, we define $K \le 32,000$ as the number of paths
for which we record the target state. Then, the sample mean $n$ conditioned
on the target measurement state is $ \langle n \rangle = \sum_{i=1} ^{K} n_i\big/K$ where we exclude the cases that we did not detect. 
Far from the topological transitions,
the option of zero detection per run (null measurement) of size $N$,
is rare, and then the conditional measurement and the theoretical predictions yield similar results. 
However, as mentioned, close to topological transitions, 
we see deviations, which are due to the finite value of $N$.
It should be noted that these deviations are a blessing in the sense that a precise measurement of a jump of $\langle n\rangle$, for a special value of $\tau$, as presented as black solid lines in Figs.~\ref{fig:exponsite} and \ref{fig:exptracking}, is non-physical, as the width of the
transition cannot be zero.
In the asymptotic theory ($N \rightarrow \infty$)
for a special value of model parameters (here $\gamma \tau$ and $\alpha$),
$\langle n \rangle$ exhibits a pointwise discontinuity
related to a jump in the winding number and the dark lines in Fig.~\ref{fig3D_Ex} become infinitely thin.

Remarkably, the simulations for the first hitting time (Figs.~\ref{fig:exponsite} and \ref{fig:exptracking}) and the detection probability (Fig.~\ref{fig3D_Ex}) agree with the results provided by the IBM quantum processor without fitting. 
This led us to the conclusion that noise is not an essential factor here. In particular, the return time $\langle n \rangle$ seems to be protected from noise for values until $N = 20$, whereas interference effects that lead to dark states are only unaffected by noise up to $N = 10$. 
The influence of noise for larger $N$ will be discussed in a forthcoming publication.

\begin{figure}\begin{center}

\includegraphics[width=0.7\linewidth]
{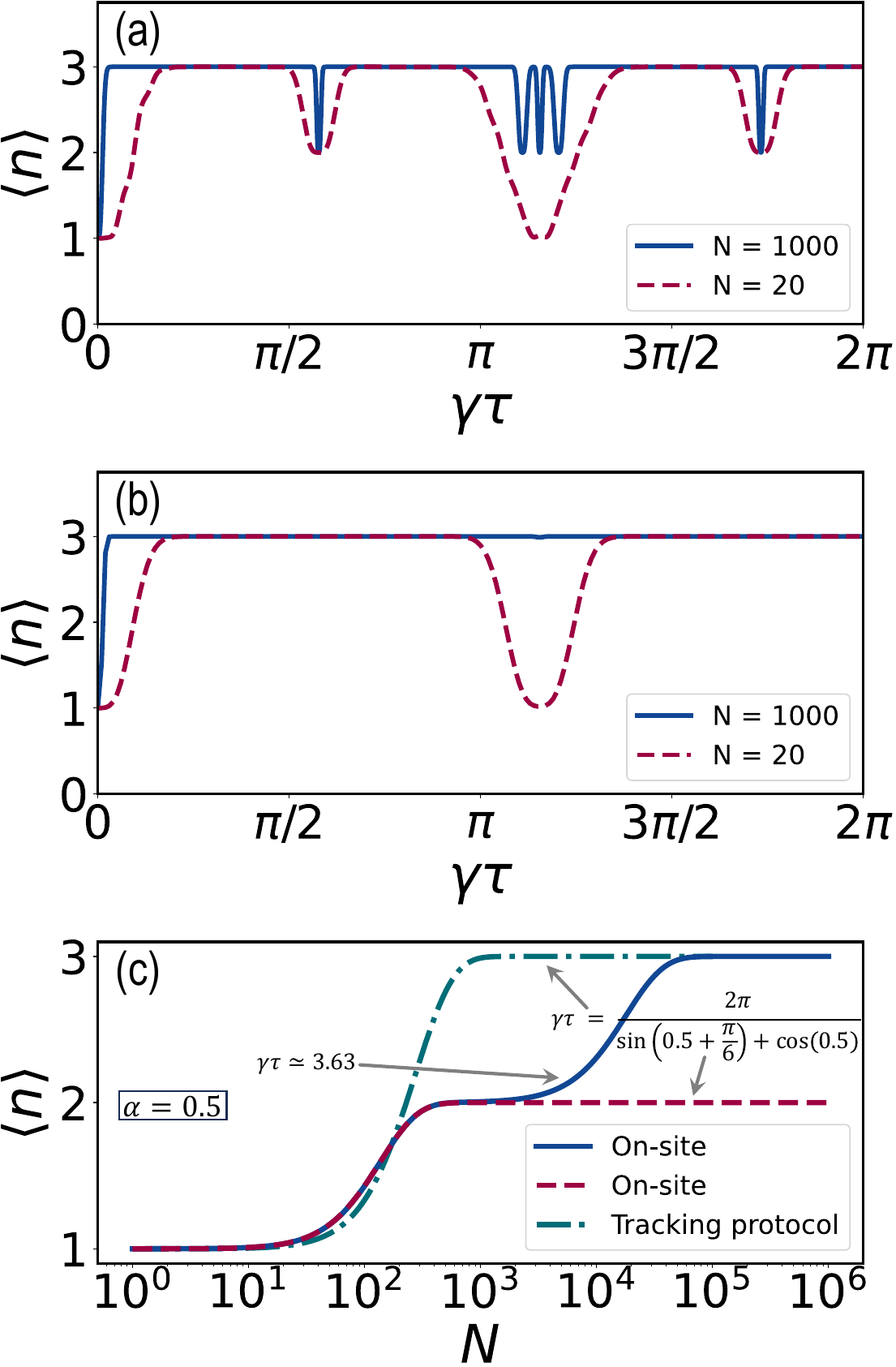}
\end{center}
\caption{ 
Simulation of the mean return
time $\langle n \rangle $ versus $\gamma \tau$ and $N$ for $\alpha = 0.5$.
(a) For $N=20$ (red dashed line) the fine structure at $\gamma \tau \approx 3.49, \ 3.63 $ and $3.78$ predicted by asymptotic theory $N \to \infty$ is washed out,
unlike the case where $N=1000$ (blue solid line). (b) At $\gamma \tau =  3.63 $ a transition from $\langle n \rangle =3$ to $\langle n \rangle =1$ and back is present for $N = 20$ (red dashed line) which is absent for $N = 1000$ (blue solid line) . 
(c) Mean $\langle n \rangle$ versus $N$ for $\alpha =0.5$ and $\gamma \tau \approx 3.63$ (blue solid line, on-site protocol), $\gamma \tau = 2 \pi/(\sin(0.5+\pi/6)+\cos(0.5))$ (green dotted-dashed line, tracking protocol) and $\gamma \tau = 2 \pi/(\sin(0.5+\pi/6)+\cos(0.5)) $ (red dashed line, on-site protocol). For an increasing number of midcircuit measurements, there is a transition from $\langle n \rangle =1$ to $\langle n \rangle =3$ (tracking protocol) that drives the system to the high-temperature limit. For the on-site protocol exactly at $\gamma \tau = 2 \pi/(\sin(0.5+\pi/6)+\cos(0.5))$ a crossover is observed from $\langle n \rangle =1$ to $\langle n \rangle =2$ and for $\gamma \tau \approx 3.63$ a crossover between three topological phases from $\langle n \rangle =1$ to $\langle n \rangle =2$ and $\langle n \rangle =3$. 
}
\label{fig:sim_20_1000}
\end{figure}
For the finite $N$ case, e.g., 
in experiments, we gain three important insights which are due to finite resolution: (i) the broadening of the transition (Figs~\ref{fig:exponsite}, \ref{fig:exptracking} and \ref{fig3D_Ex}), (ii) additional dips and plateaus [Figs.~\ref{fig:exponsite}~(e) and \ref{fig:exptracking}~(d) and (e)] and a chirality effect of the total detection probabilities (Fig.~\ref{fig:chiral}). 

We have already mentioned that for the on-site measurement and $\alpha=0.5$, three dips found by theory are not found in the experiment; see Fig.~\ref{fig:exponsite} (d).
Furthermore, for the tracking protocol, a wide resonance is found in the experiment which is absent in theory [Fig.~\ref{fig:exptracking} (d)]. 
We claim that this is due to finte $N$ effects. 
To study this, we consider the mean return time, obtained from simulations, for various $N$.
As shown in Figs.~\ref{fig:sim_20_1000} (a) and (b), for $N=1000$ the resonances reveal themselves. However, that resolution is not yet experimentally available.
The effects of finite resolution lead to the possibility of a crossover between different topological winding numbers [Fig.~\ref{fig:sim_20_1000} (c)] from $\langle n \rangle =1$ to $\langle n \rangle =2$ and $\langle n \rangle =3$ with an increasing number of measurements $N$, i.e., near the special point $(\gamma \tau, \alpha) = (2 \pi/\sqrt{3}, \pi/6)$ for the on-site protocol, which can be interpreted as a metastable topological effect. For the tracking protocol, the crossover from $\langle n \rangle =1$ to $\langle n \rangle =3$ can be understood as a partial thermalization for a small number of measurements.

In Appendices \ref{appendix:onsite} and \ref{appendix:Didi}
the eigenvalue spectrum of the survival operator and the stochastic matrix are analyzed for the on-site and tracking protocol, respectively.
$\langle n \rangle$ is discontinuous at special points
when the absolute value of the eigenvalue is one. For $N =20$ resonances and plateaus are also present when the absolute values of the eigenvalues are almost one, leading to metastable topological effects and partial thermalization. 
This, as mentioned,
is because the observation is for a limited duration, i.e., $N \tau$ is
finite, and because of the slow decay of the null measurement probability. 
Both are discussed in the following two subsections.

\subsection{Broadening effect}
	We present an exact solution to the problem related to our experimental setup for both on-site and tracking measurement protocols. Previously, the broadening effect was studied in more generality for the on-site measurements, although the asymptotic behavior of large $N$ was studied there \cite{yin2024restart}.
  In the following we will discuss the broadening of the return time $\langle n \rangle$ for zero magnetic flux in detail.
 \subsubsection{On-site protocol (broadening)}
We first consider the broadening effect of the resonances of $\langle n \rangle$ using the on-site measurement protocol
and $\alpha=0$.
Let $F_n$ be the probability of finding the target for the first time
at the $n-$th measurement.
The case under study is exactly solvable
\cite{Yin2019,PhysRevResearch.5.033089}. Specifically, with the eigenvalues of Eq. (\ref{eigenvalues1}) we obtain
\begin{equation}
F_n = \left\{
\begin{array}{c c}
|z|^2 \  & n =1 \\
\ & \ \\
\left(1 - |z|^2\right)^2 |z|^{2 (n-2)} \ & n \ge 2,
\end{array}
\right.
\label{Fnz2}
\end{equation}
where $|z|^2= 5/9 + 4/9 \cos( 3 \gamma \tau)$.
Here, we stick to the notation in~\cite{Yin2019}, where in general, $z$ stands for the zeros
of the generating function. 
When $|z|^2 = 1$, we have $F_1=1$ and then $\langle n \rangle=1$.
Otherwise, $\langle n \rangle = \sum_{n=1} ^\infty n F_n= 2$,
which is a behavior that was presented in Fig.~\ref{fig:exponsite} (c). 
 Clearly, when $|z|^2$ is slightly less than one, so is $F_1$,
yet from the recurrence theorem \cite{Gruenbaum2013,Grunbaum2014} we have $\lim_{N \rightarrow \infty}\langle n (N) \rangle =2$ implying a very slow decay of 
$F_n \propto |z|^{2(n-2)}$ 
for $n>1$. 
Physically, close to $|z|^2=1$ we usually detect the system with the first click,
but in rare events, we click no. Then the wave function does not overlap with the detected state, and recording a yes becomes unlikely. On average we find that the return click yes after
$\langle n \rangle =2$
attempts which is remarkable since $F_1$ is nearly one.

 In a finite-time experiment, we find 
 $\langle n \rangle$ [Eq.~(\ref{eq:condn})]
using Eq.~(\ref{Fnz2})
\begin{equation}\label{eq8}
\langle n \rangle = { 2 - |z|^{2(N-1)} \left[ 1 + N (1 - |z|^2) \right] \over
1 - |z|^{2(N-1)}  \left( 1 - |z|^2\right) }.
\end{equation}
We will study this expression in the vicinity of the revival
times, namely when $|z|^2 \to 1$, where we also include the Zeno limit in this category. 
Note that if $|z|^2 =1$ we have $\langle n (N) \rangle = 1$ for any $N$.

We note that the expressions in Eqs.~(\ref{Fnz2}) and (\ref{eq8}) have the same form for a two-level
system, except for a different function of $|z|^2$~\cite{PhysRevResearch.5.033089}. 
This indicates that for $\alpha=0$ the  three-site model represents an effective 
two-level system.

\begin{figure}\begin{center}

\includegraphics[width=0.6\linewidth]
{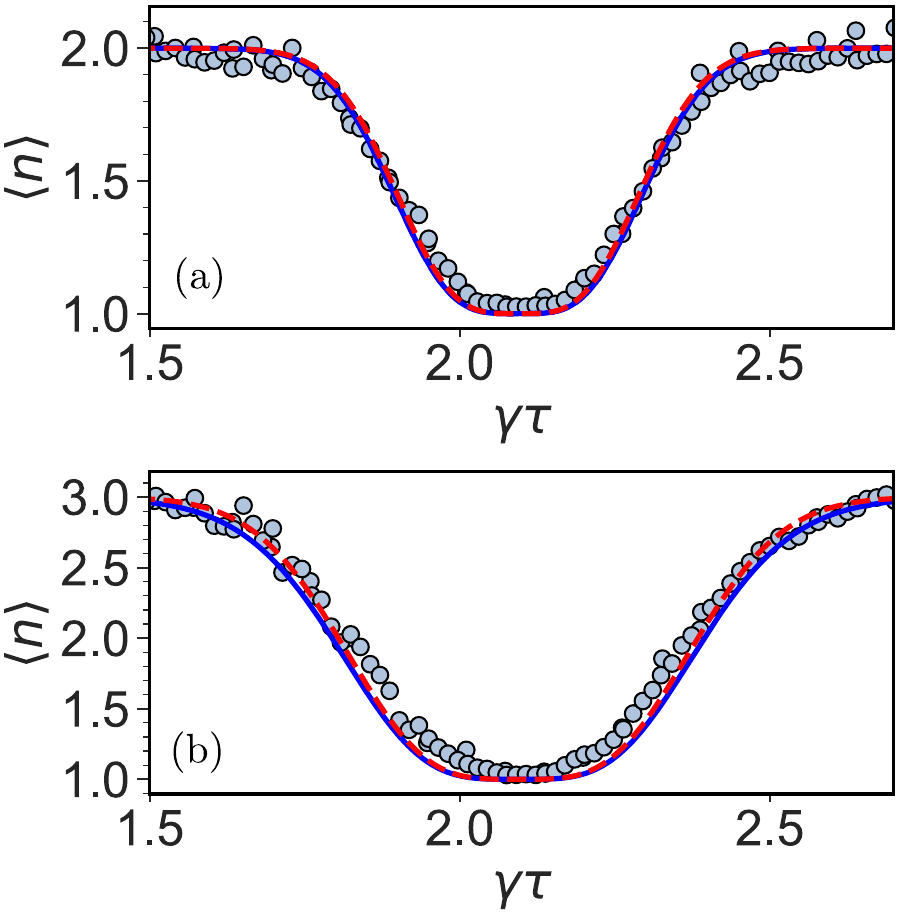}
\end{center}
\caption{ (a) The transition from $\langle n \rangle=2$ to $\langle n \rangle=1$
and back is widened due to the finite time of the experiment. We compare IBM quantum processor results (light blue circles) and theory  $\langle n \rangle = 2 - \exp(- b) (1 + b)$
with
 $b= N (2/9) (3 \gamma \tau - 2 \pi)^2$  for $N=20$ (red dashed line) and simulation (blue solid line).
 (b) The transition from $\langle n \rangle=3$ to $\langle n \rangle=1$
and back is widened due to the finite time of the experiment. We compare IBM quantum processor results (light blue circles) and theory  $\langle n \rangle = 3 - 2 \exp(- b/2) (1 + b/2)$
with
 $b= N (2/9) (3 \gamma \tau - 2 \pi)^2$  for $N=20$ (red dashed line) and simulation (blue solid line).
}
\label{fig:br}
\end{figure}

Close to transitions we use the small parameter $\epsilon^2 = 1 - |z|^2$. For the model under study, $\epsilon^2 =(2/9)(3 \gamma \tau - 2 \pi k)^2$ where $k=0,1,...$ is the index for the $k$-th transition of $\langle n \rangle$. For example, $k=0$ is the Zeno limit which corresponds to the first dip from the left
in the upper panel in Fig.~\ref{fig:exponsite} (a).
Taking the limit of large $N$ and $\epsilon^2$ to be small, we find
\begin{equation}\label{eq:br1}
\langle n \rangle = 2 - \exp( - N \epsilon^2) \left(1 + N \epsilon^2\right).
\end{equation} 
Recall that close to the transition point 
$\langle n \rangle = 1$ for large $N$. Here by transition we mean a transition for $\langle n \rangle=2$ to $\langle n \rangle=1$ as we vary gamma tau across a resonance. 

On the other hand,
if we choose a control parameter like $\gamma \tau$ to be far from the transition, then $\langle n \rangle \approx  2$, since $N$ is large. 

\subsubsection{Tracking protocol (broadening)}
Now let us consider the broadening effect of the resonances of $\langle n \rangle$ using the tracking protocol
and $\alpha=0$.
The probability of finding the quantum walker the first time is
\begin{equation}
F_n = \left\{
\begin{array}{c c}
|z|^2 \  & n =1 \\
\ & \ \\
2|\eta|^4 \left(|\eta|^2 + |z|^2\right)^{n-2}  \ & n \ge 2,
\end{array}
\right.
\label{eq:Fnztrack}
\end{equation}
where $|z|^2= 5/9 + 4/9 \cos( 3 \gamma \tau)$ and $|\eta|^2 = 2/9-2/9 \cos( 3 \gamma \tau)$. With $|\xi|^2 = |z|^2+|\eta|^2$
 the mean number is:
\begin{eqnarray}
\langle n \rangle &=& \frac{2 |\eta|^4 \left( |\xi|^4 - 2|\xi|^2 + |\xi |^{2N}  \left( -N |\xi|^2 + N + 1 \right) \right) }{\left( |\xi|^2 - 1 \right) \left( 2|\eta|^4 \left( |\xi|^2 - |\xi |^{2N}  \right) - |\xi|^2 |z|^2 \left( |\xi|^2 - 1 \right) \right)} \nonumber\\
&-&\frac{ |\xi|^2 |z|^2 \left( |\xi|^2 - 1 \right)}{ \left( 2|\eta|^4 \left( |\xi|^2 - |\xi |^{2N}  \right) - |\xi|^2 |z|^2 \left( |\xi|^2 - 1 \right) \right)}
\end{eqnarray}
For small $\epsilon^2 =(2/9)(3 \gamma \tau - 2 \pi k)^2$ close to the transition we get an expression for the tracking case:
\begin{eqnarray}
\label{eq:br2}
    \langle n \rangle \approx 3 - 2 \exp(-N\epsilon^2 /2 ) (1+N  \epsilon
^2/2 )  
\end{eqnarray}
Close to the transition $\langle n \rangle = 1$ for any large enough $N$. If we choose a control parameter like $\gamma \tau$ to be
far from the transition, then $\langle n \rangle = 3$.
Fig.~\ref{fig:br} compares the approximations for the on-site and tracking protocol in Eq.~\ref{eq:br1} and Eq.~\ref{eq:br2} with the simulation and experimental data.

\begin{figure}\begin{center}
\includegraphics[width=0.8\linewidth]
{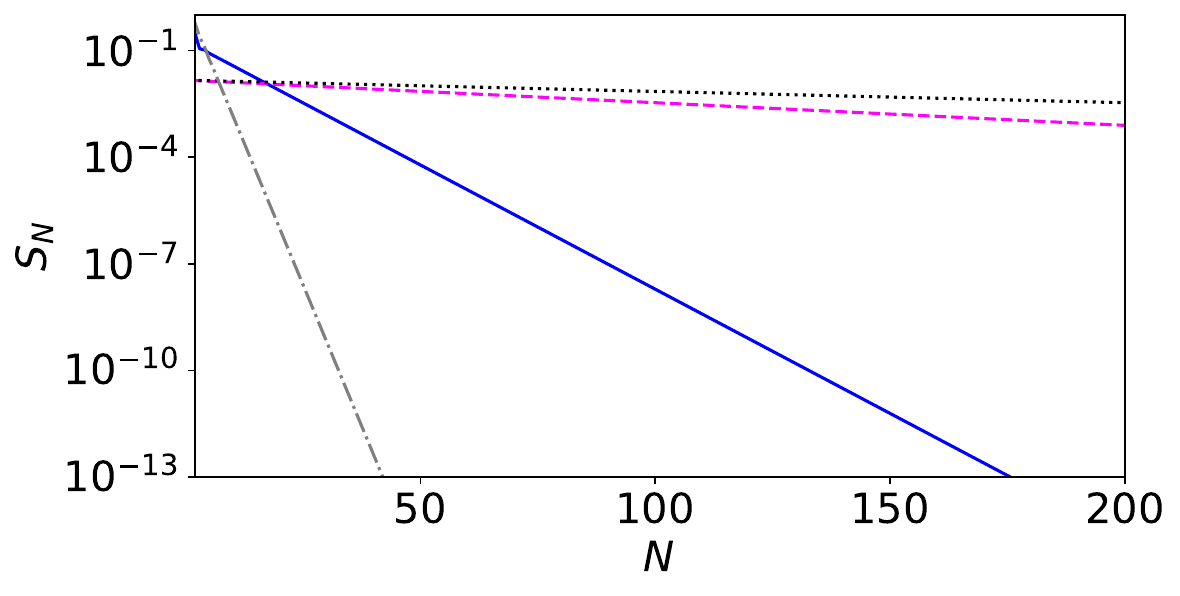}
\end{center}
\caption{ The null measurement  probability versus $N$ for $\alpha =0.5$ for the tracking and on-site protocol.
 The parameters for the tracking protocol are $\gamma \tau = 3.63$ and $\alpha =0.5$ (magenta, dashed), and $\gamma \tau = 2$ and $\alpha =0.5$ (grey, dashed dotted), and for the on-site protocol $\gamma \tau = 3.63$ and $\alpha =0.5$ (black, dotted), and $\gamma \tau = 2$ and $\alpha =0.5$ (blue, solid). The null measurement  rate slows down considerably near
special sampling rates at $(\gamma \tau, \alpha) = (3.63, 0.5)$ for both protocols.}
\label{fig:surv}
\end{figure}
\subsection{Slow decay of the null measurement  probability}

To explain the experimental findings of an additional dip (transition) in the mean first return time $\langle n \rangle$ at $\alpha = 0.5$, $\gamma \tau = 3.63$ for both protocols [see Figs.~\ref{fig:exponsite} (d) and (e) and \ref{fig:exptracking} (d) and (e)] as well as the plateau near $\alpha = 0.5$ for the on-site protocol in Fig.~\ref{fig:exponsite} (e), which are absent in the theory for $N \rightarrow \infty$, 
we investigate the probability of null measurement, i.e., the probability that the quantum walker is not detected.
The null measurement probability is defined as
\begin{eqnarray}
    S_N = 1- \sum_{n=1}^N F_n.
\end{eqnarray}
It is zero in the limit $N \rightarrow \infty$ for the return problem \cite{Gruenbaum2013,Grunbaum2014,Friedman2017}.
Near special points where the three phase factors merge,
the null measurement probabilities decay only very slowly to zero for increasing $N$, 
leading to the transition of $\langle n \rangle$ to 3 (tracking protocol) and 2 (on-site protocol) only for large $N>1000$ for $\gamma \tau = 2 \pi/(\sin(0.5+\pi/6)+\cos(0.5)) $ [Fig.~\ref{fig:sim_20_1000}~(c)]. If $\gamma \tau \approx 3.63$, $\langle n \rangle$ shows a staircase behavior.
The system is only partially thermalized until, for a larger number of measurements, 
the high-temperature limit is reached.

The probability of null measurement decreases exponentially to zero at non-special sampling rates, whereas the decay rate slows down considerably at special sampling rates.
At the special point $(\gamma \tau, \alpha) = (2 \pi/\sqrt{3}, \pi/6)$, 
the mean first return time
$\langle n \rangle = 1$, since the three phase factors match. This is also approximately the case near the special point $(\gamma \tau, \alpha) = (3.63, 0.5)$ due to finite resolution. The null measurement  probability is displayed in Fig.~\ref{fig:surv} for the tracking and on-site protocol. This behavior is very similar to the Zeno effect at $\tau \approx 0$ and is discussed in detail in the Appendices~\ref{appendix:onsite}
and \ref{appendix:Didi} by analyzing the eigenvalues of the survival operator (on-site protocol) and the stochastic matrix (tracking protocol), respectively. 

\section{Summary and Conclusion\label{Con}}

We experimentally and theoretically investigated the monitored evolution of a quantum walker on a ring represented
by a directed triangle graph with complex edge weights -- a monitored chiral quantum walk.
For this purpose, we used the capabilities of midcircuit measurements on IBM quantum devices. We were able to accurately confirm the predictions for the hitting times and detection probabilities of the general theory for a finite number of measurements. Remarkably, we utilize only dynamical decoupling as an error suppression scheme and no other error mitigation methods. 

Hitting times and detection probabilities are closely related to the number of distinct eigenvalues (phase factors) of the unitary time-evolution operator $U = \exp (-i H \tau)$ of the model Hamiltonian $H$ during the measurement-free evolution time $\tau$ shown in Fig.~\ref{fig:phases}.
To investigate them, we swept through the parameters of the phase-factor matching diagram and performed two different stroboscopic measurement protocols.
 In the first protocol, readout was performed only at the target site. The second protocol, known as the tracking protocol, involves measurements at each site along the ring.
The main experimental result
is that there are clear differences between both protocols, a feature not known for
classical random walks. 
The onsite protocol shows quantum behavior. The mean return time is quantized and linked to a winding number. Moreover, the winding
number of the return problem is equal to the dimensionality of the available Hilbert space. This was originally
discovered in Ref.~\cite{Gruenbaum2013,Grunbaum2014} for periodic measurements
and later found also for measurements at random times \cite{KesslerRand}.
Since the dimensionality of the available Hilbert space changes when degenerate states occur, the winding number as well as the mean return time can jump between discrete values.

The tracking protocol exhibits quantum features when all the phase factors match, but otherwise shows classical behaviors. 
The results show that the postulates of measurement theory as well as the presence of a phase utilising the onsite protocol could be confirmed on the IBM quantum computer, a non-trivial fact since $N \sim 20$ repeated noisy measurements are utilized.

We showed that finite resolution effects, which are clearly part of any experimental study, are a key feature of the hitting time statistics.
 This leads to a broadening of the mean return time and the detection probability because the detection is imperfect near special points, where two or three phase factors merge, due to finite $N \tau$.
The finite resolution and the broadening effect are accompanied by a slow relaxation of the null measurement probability. 
The results show not only a broadening of the topological transitions but also metastable topological [see Fig.~\ref{fig:exponsite}~(e) and see Fig.~\ref{fig:exptracking}~(e)] and chirality effects (see Fig.~\ref{fig:chiral}), which disappear for the theory in the asymptotic limit. 

We derived an analytical expression for the broadening of the $\langle n \rangle = 2$ to $\langle n \rangle = 1$ and back transition (on-site) and the $\langle n \rangle = 3$ to $\langle n \rangle = 1$ and back transition (tracking) for $\alpha = 0$ as a function of $N$. For both protocols, the resonances narrow as $N$ increases.
Furthermore, the additional resonances near the special points (but not exactly on them) disappear
for growing $N$. The mean return time $\langle n \rangle $
 shows a crossover between different topological phases, which is revealed, e.g., as the staircase behavior for the on-site protocol [Fig.~\ref{fig:sim_20_1000}~(c)]. A similar effect was found theoretical for a perturbed ring \cite{Wang}.
As mentioned, the tracking protocol of the monitored quantum walk has more of the character of a classical random walk.
Because of the finite resolution, quantum effects are present close to the revival time. This can be seen as partial thermalization \cite{Santini_2023} during a quantum-to-classical crossover. Only for a large number of measurements the classical limit and maximum $\langle n \rangle$ is reached, except for special $\gamma \tau$ and $\alpha$.

There are several future directions in which our work can be extended.
Here, we focus on small systems.
More generally, one could scale up
the process by increasing the number of measurements and/or increasing the size of the system.
One could then ask whether the quantum computer exhibits a transition to a more classical behavior, for example, the elimination of the topological effect found for the mean return
time or how a partial information about the trajectories by measuring multiple but not all sites change the topological effect.
Furthermore, one could explore the possibility of measuring the readout rate-controlled oscillatory behavior of the energy expectation values corresponding to a certain trajectory of the monitored dynamics.
In a future publication, we will show that noise in the system enables transitions to theoretically forbidden
states, an effect that becomes particularly important as we increase $N$.


\section{Acknowledgments}
We acknowledge the use of IBM Quantum services for this work. The views expressed are those of the authors and do not reflect the official policy or position of IBM or the IBM Quantum team. In this paper, we used the IBM {\it Sherbrooke} processor which is an IBM Quantum Eagle  Processor.
Q.W., S.R. and R.Y. acknowledge the use of the IBM Quantum Experience and the IBMQ-research program.
We acknowledge helpful discussions with David Kessler. 
The support of the Israel Science Foundation grant 1614/21 is acknowledged.

%


\section{Author Contribution}

E.B. and S.T. conceived and designed the project. 
S.T., S.R., and Q.W. ran hardware, numerical experiments, and classical simulations. 
All authors discussed the theory and the results. E.B. and S.T. drafted the manuscript with input from all authors.

\appendix
\begin{figure}\begin{center}
\includegraphics[width=1\linewidth]{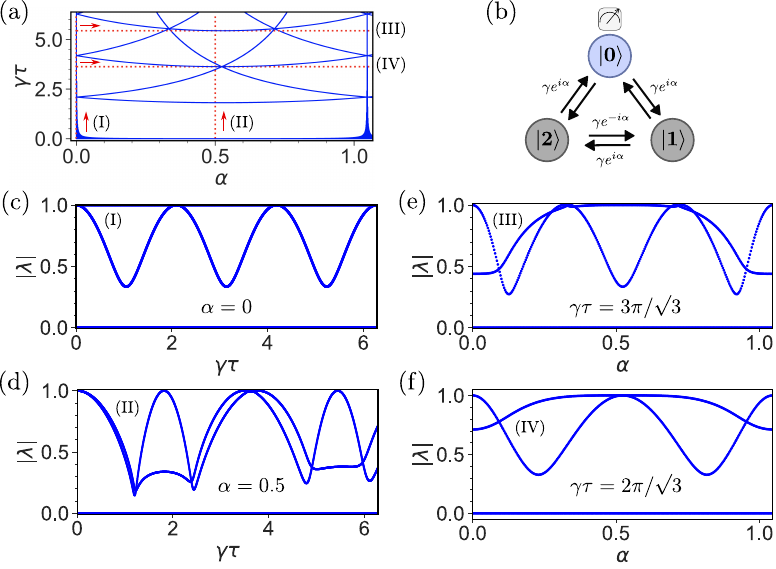}
\end{center}
\caption{
(a) Phase factor matching diagram and paths through the parameter space labeled (I), (II), (III), and (IV) which are indicated with red dashed lines and correspond to (c), (d), (e), and (f), respectively.
The eigenvalues close to one lead to additional dips and plateaus in the mean return time, whereas the eigenvalues exactly equal to one indicate transitions for the theory $N \rightarrow \infty$.
(b) Illustration of the on-site measurement protocol. (c) - (f) Eigenvalue spectrum of the survival operator for the same parameters used for the mean return time in Fig.~\ref{fig:exponsite}. 
(c) For $\alpha = 0$ the eigenvalues are one for $\gamma \tau =0,  2 \pi/3 , 4 \pi/3$ and $6 \pi/3$, leading to transitions in the mean return time [path (I)]. (d) For $\alpha = 0.5$ the eigenvalues are one for $\gamma \tau \approx 0, 1.82, 3.50, 3.62, 3.78, 5.44$. The three eigenvalues $|\lambda| = 1$ for $ \gamma \tau \approx 3.50, 3.62, 3.78$ are very close, so almost three phase factors match leading to a transition to one of the mean return times [path (II)].
In (e) [path (III)] and (f) [path (IV)] $|\lambda|$ is close to one between $\alpha \approx 0.3$ and $\alpha \approx 0.7$ leading to a plateau for $N = 20$. }
\label{fig:eigenonsite}
\end{figure}

\begin{figure}\begin{center}
\includegraphics[width=1\linewidth]{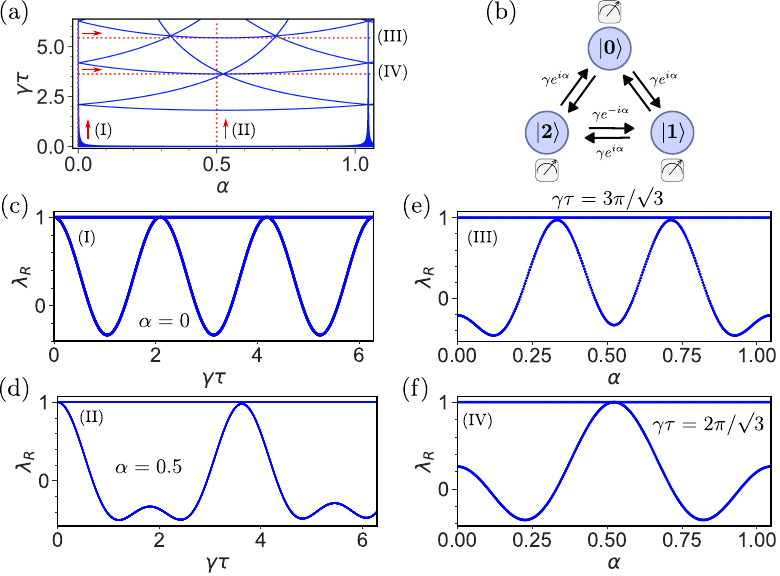}
\end{center}
\caption{ (a) Phase factor matching diagram and paths through the parameter space labeled (I), (II), (III), and (IV) which are indicated with red dashed lines and correspond to (c), (d), (e), and (f), respectively.
 (b) Illustration of the tracking protocol.
 (c) - (f) The eigenvalues of $G_{\rm tar}$ for the same parameters as the mean return time in Fig.~\ref{fig:exptracking}.
Two nearly degenerate eigenvalues close to one lead to additional dips in the mean return time, while two degenerate eigenvalues exactly at one indicate transitions for the theory $N \rightarrow \infty$.
(c) For $\alpha = 0$ the eigenvalues are degenerate, and one for $\gamma \tau = 0,  2 \pi/3 , 4 \pi/3$ and $6 \pi/3$ which leads to transitions of the mean return time [path (I)].
(d)  For $\alpha = 0$ and $\gamma \tau = 3.63$ the eigenvalues are almost degenerate and one which leads to a transition for $N = 20$ due to finite resolution effects [path (II)]. (e) For $\gamma \tau = 3 \pi/\sqrt{3}$ the eigenvalues are almost degenerate and one for $\alpha \approx  0.3 $ and $\alpha \approx  0.7 $ [path (III)].
 (f)  For $\gamma \tau = 2 \pi/\sqrt{3}$ the eigenvalues are degenerate and one for $\gamma \tau = \pi/6 $ indicating a transition for all $N$ [path (IV)].}
\label{fig:lambdaalp05}
\end{figure}
\section{\label{phase}phase factor matching diagram and eigenstates}
This appendix provides the analytical solutions of the phase factor matching equations and the eigenstates. 
As mentioned in the text the eigenvalues of 
$$H = -\gamma \ e^{i \alpha } \left( \ket{\bf{0}} \bra{\bf{1}} + \ket{\bf{1}} \bra{\bf{2}}+\ket{\bf{2}} \bra{\bf{0}} \right) + {\rm h.c.}$$ are
$$E_k= - 2 \gamma \cos\left( {2 \pi k \over 3} + 
\alpha\right),  \ k=0,1,2 \ .$$ Its eigenvectors are for finite $\alpha$:
$$ \ket{E_0} = \frac{1}{\sqrt{3}} \left(\ket{\bf{0}}+\ket{\bf{1}}+\ket{\bf{2}}\right),$$
$$ \ket{E_1} = \frac{1}{2} \left(\left(\frac{-1}{\sqrt{3}}-i \right)\ket{\bf{0}}+\left(\frac{-1}{\sqrt{3}}+i\right)\ket{\bf{1}}+\frac{2}{\sqrt{3}}\ket{\bf{2}}\right)$$
and 
$$ \ket{E_2} = \frac{1}{2} \left(\left(\frac{-1}{\sqrt{3}}+i \right)\ket{\bf{0}}+\left(\frac{-1}{\sqrt{3}}-i\right)\ket{\bf{1}}+\frac{2}{\sqrt{3}}\ket{\bf{2}}\right)$$
The eigenstates for $\alpha = 0$
are:
$$ \ket{E_0} = \frac{1}{\sqrt{3}} \left(\ket{\bf{0}}+\ket{\bf{1}}+\ket{\bf{2}}\right)$$
and for the degenerate eigenvalue $E_1=E_2$:
$$ \ket{E_1} = \frac{1}{\sqrt{2}} \left(-\ket{\bf{0}}+\ket{\bf{1}}\right)$$
and 
$$ \ket{E_2} = \frac{1}{\sqrt{2}} \left(-\ket{\bf{0}}+\ket{\bf{2}}\right)$$

The solutions of the following equations (matching of phase factors):
\begin{eqnarray}
 \exp (-i E_k \tau) = \exp(-i E_l \tau) \ ,  (k \neq l), \ k,l = 0,1,2   
\end{eqnarray}
are for $k=1$ and $l=2$:
\begin{eqnarray}
    \gamma \tau = \frac{2n\pi}{\sqrt{3} \sin(\alpha)}
\end{eqnarray}
and 
\begin{eqnarray}
    \gamma \tau = -\frac{(2n+1)\pi}{\sqrt{3} \sin(\alpha)}
\end{eqnarray}
for $k=1$ and $l=0$:
\begin{eqnarray}
    \gamma \tau =\frac{n\pi}{\sin(\alpha+\pi/6)+\cos(\alpha)}
\end{eqnarray}
for $k=2$ and $l=0$:
\begin{eqnarray}
    \gamma \tau =\frac{n\pi}{\sin(-\alpha+\pi/6)+\cos(\alpha)}
\end{eqnarray}
for $n \in \mathbb{Z}$.
 
\section{\label{appendix:onsite} Details of the on-site protocol}
The probability $F_n$ to find the quantum walker for the first time is for the on-site protocol \cite{Friedman2017}:
\begin{eqnarray}
    F_n = |\langle \psi_{tar} | \left(U (\bf{I}-| \psi_{tar}  \rangle\langle \psi_{tar} | \right))^{n-1} U | \psi_{in}  \rangle|^2
\end{eqnarray}
where $\bf{I}$ is the identity matrix, $|\psi_{tar} \rangle$ is the target site and $|\psi_{in} \rangle$ is the initial site. 
To explain the experimental findings of a transition from $\langle n \rangle = 3$ to $\langle n \rangle = 1$ for $\alpha = 0.5$, $\gamma \tau = 3.63$ and $N=20$ in Fig.~\ref{fig:exponsite}~(c) and in the simulation Fig.~\ref{fig:sim_20_1000}~(a)
which is changes to three close transitions from $\langle n \rangle = 3$ to $\langle n \rangle = 2$ (on-site protocol) for $N > 1000$ we analyse the survival operator for the on-site protocol.

The parameter regime is indicated by a dashed vertical line in Fig.~\ref{fig:eigenonsite} (a) at $\alpha = 0.5$. The line crosses five phase lines where two phase factors merge. Three of them are very close. For $N$ very large in the on-site protocol, five transitions from $\langle n \rangle = 3$ to $\langle n \rangle = 2$ are seen. 

What is the effect of the nearby special point ($\alpha = \pi/6$, $\gamma \tau = 2\pi/\sqrt{3}$) where the three phase factors merge? Does it lead to a slow relaxation of the null measurement probability and therefore to the aforementioned transition which is present for $N = 20$ measurements?

The survival operator for the on-site protocol is defined as
\begin{eqnarray}
    S(\tau) = \left({\bf{I}}-\ket{\psi_{\rm tar}}\bra{\psi_{\rm tar}}\right) U
\end{eqnarray}

The eigenvalues of the survival operator are key to understanding the processes and to answering the questions at the beginning of the section.
The absolute values of two eigenvalues $|\lambda_1| \approx |\lambda_2|  $ are close to one for $\gamma \tau = 3.63$ [see Fig.~\ref{fig:eigenonsite} (d)].  
The plateau in Fig.~\ref{fig:exponsite}~(e) coincides with the eigenvalue $|\lambda| \approx 1$ between $\alpha = 0.3$
and $\alpha = 0.7$.

\section{\label{appendix:Didi} Details of the tracking protocol}
For the time evolution that takes into account unitary evolution and global measurements, we define the Markov or stochastic matrix $G$ \cite{Didi}:
\begin{eqnarray} \label{eq:stoch}
    G = \sum_{x,x' \in X}\left| \bra{x} U \ket{x'}\right|^2 \ket{x}\bra{x'} ,
\end{eqnarray}
where $X=\{\ket{\bf{0}},\ket{\bf{1}},\ket{\bf{2}}\}$ and $\left| \bra{x} U \ket{x'}\right|^2 $ are the transition probabilities.
The probability $F_n$ of finding the quantum walker for the first time is then:
\begin{eqnarray}
    F_n = \langle \psi_{\rm tar} | \left(G ({\bf{I}}-| \psi_{\rm tar}  \rangle\langle \psi_{\rm tar} |) \right)^{n-1} G | \psi_{\rm in}  \rangle,
\end{eqnarray}
For $\alpha = 0$ the eigenvalues of $G$ are real and the absolute value is less than or equal to one. $\langle n \rangle$ is discontinuous at special points when the eigenvalue $\lambda = 1$ becomes degenerate. For a finite magnetic flux, the eigenvalues except one ($\lambda = 1$) become complex and appear as conjugated complex pairs. There are special points for finite $\alpha = \pi/6,\pi/3,\pi/2,...,$ where for certain values of $\gamma \tau$ the eigenvalues of $G$ are degenerate and one (see Fig.~\ref{fig:lambdaalp05}).
Therefore, no special points appear at $\alpha = 0.5$ and $\langle n \rangle = 3$ for all $\gamma \tau$ except in the Zeno regime for $\gamma \tau = 0$. In Fig.~\ref{fig:sim_20_1000} (b) $\langle n \rangle$ with a finite magnetic flux is shown. For $N= 1000$ no transitions from 3 to 1 and back are present (except in the Zeno limit) while for $N =20$ the transitions are clearly present due to finite resolution. Following Ref.~\cite{Didi}
we define:

\begin{eqnarray}
    G_{\rm tar} = G \left({\bf{I}}-\ket{\psi_{\rm tar}}\bra{\psi_{\rm tar}}\right)
\end{eqnarray}
$G \ket{\psi_{\rm in}}$ is written in the eigenbasis of $G_{\rm tar}$. 
The left and right eigenstates are defined in the usual way: $G_{\rm tar} \ket{\mu_R} = \mu \ket{\mu_R} $ and $\bra{\mu_L} G_{\rm tar} = \mu \bra{\mu_L} $ where $\mu$ is the eigenvalue.
The null measurement probability (probability that the quantum walker is not detected) is then given by:
\begin{eqnarray}
    S_N = \sum_{\mu} \frac{e^{N \ln (\mu)}}{1-\mu} \frac{\bra{\mu_L} G \ket{\psi_{\rm in}}}{\braket{\mu_L|\mu_R}} \braket{\psi_{\rm tar}|\mu_R}
\end{eqnarray}
The null measurement  probability is characterized by  $\mu $ as well as the overlap between $\ket{\mu_R}$ and the target state $\ket{\psi_{\rm tar}}$. For the special point $(\pi/6,2\pi/\sqrt{3})$ the overlap is zero, and the eigenvalue is $\mu = 1$. Near the special point $(0.5,3.63)$ there is a finite overlap and the eigenvalue is close to one, leading to a very slow relaxation of the null measurement probability. Therefore, $\langle n \rangle = 1$ for $N = 20$.   
For an increasing number of $N$, the system experiences a transition from $\langle n \rangle =1$ to $\langle n \rangle =3$ driving the system into the high-temperature limit. Thus, for $ N= 20$ we consider that the system is still in the quantum regime.

\section{Simulated detection probability \label{app:sim}}
\begin{figure}
\begin{center}
\includegraphics[width=1\linewidth]{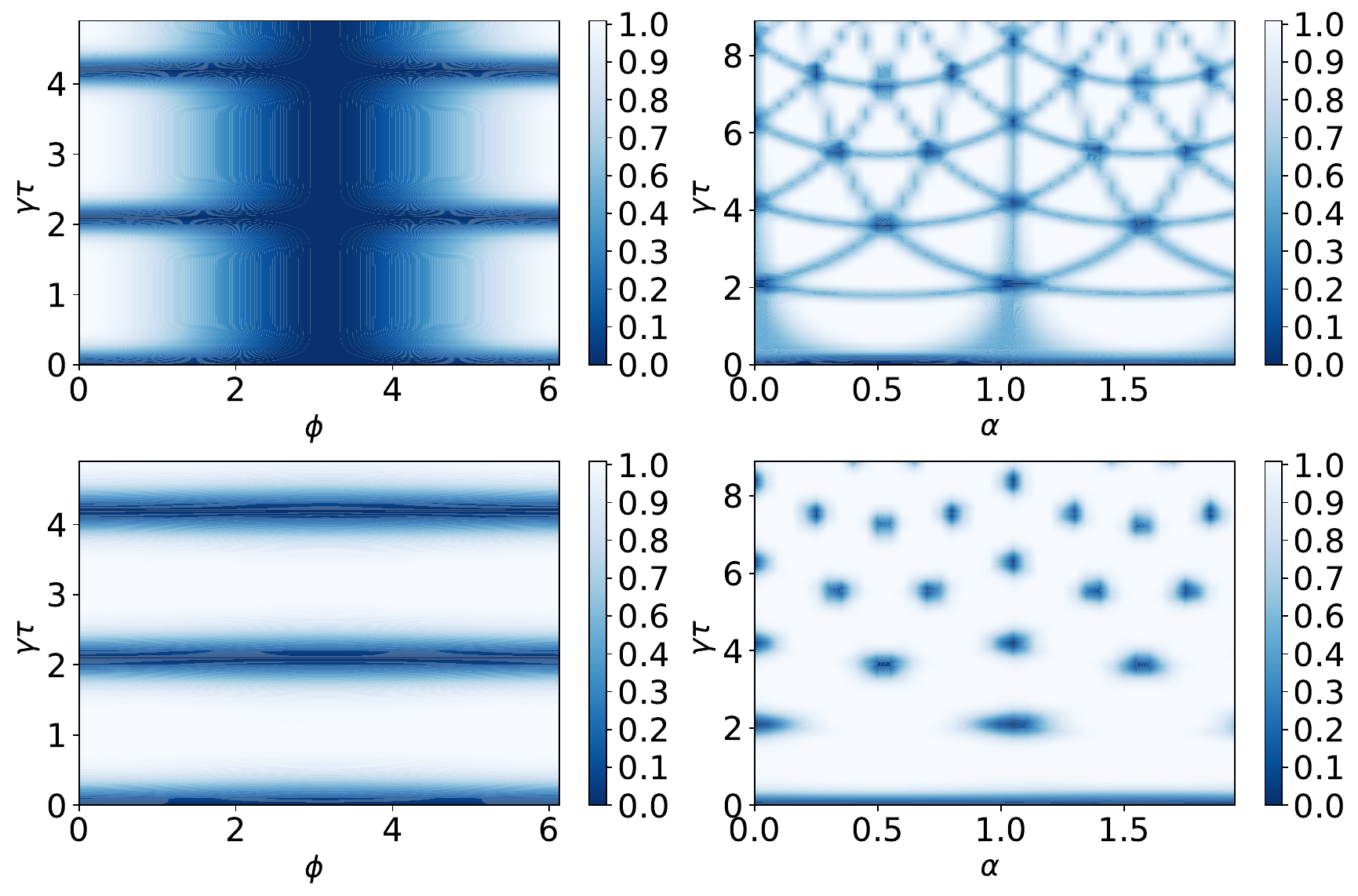}
\end{center}
\caption{ The same as Fig.~\ref{fig3D_Ex} but as a simulation. We observe that the experiment and the simulations agree very well for $N = 10$.}
\label{fig:3Dsim}
\end{figure}
In Fig.~\ref{fig:3Dsim} we present the simulation of the detection probability which is in a very good agreement with the experimental data shown in Fig.~\ref{fig3D_Ex}.

\end{document}